\documentclass[12pt,aps,superscriptaddress,a4paper]{revtex4}
\usepackage{graphicx}
\usepackage{amsmath}
\usepackage{amssymb}
\usepackage{enumerate, color}
\usepackage{subfigure}
\usepackage{tabularx}
\usepackage{epstopdf}
\usepackage{lipsum}
\usepackage{slashed}
\usepackage{tensor}
\usepackage[colorlinks=true, pdfstartview=FitV, linkcolor=blue, citecolor=red, urlcolor=black, breaklinks=true]{hyperref}
\newcommand{\be}{\begin{equation}}
\newcommand{\ee}{\end{equation}}
\newcommand{\ben}{\begin{eqnarray}}
\newcommand{\een}{\end{eqnarray}}
\newcommand{\bes}{\begin{subequations}}
\newcommand{\ees}{\end{subequations}}
\def\bal#1\eal{\begin{align}#1\end{align}}

\newcommand{\LL}{{\cal L}}

\newcommand{\LX}{{\cal L}_X}

\newcommand{\LXX}{{\cal L}_{XX}}

\newcommand{\vphi}{{\varphi}}
\newcommand{\LY}{{\mathcal{L}_Y}}
\begin{document}

\title{First Order Formalism for Generalized Vortices}
\author{D. Bazeia}\email{bazeia@fisica.ufpb.br}\affiliation{Departamento de F\'\i sica, Universidade Federal da Para\'\i ba, 58051-970 Jo\~ao Pessoa PB, Brazil} 
\author{L. Losano}\email{losano@fisica.ufpb.br}\affiliation{Departamento de F\'\i sica, Universidade Federal da Para\'\i ba, 58051-970 Jo\~ao Pessoa PB, Brazil}
\author{M.A. Marques}\email{mam@fisica.ufpb.br}\affiliation{Departamento de F\'\i sica, Universidade Federal da Para\'\i ba, 58051-970 Jo\~ao Pessoa PB, Brazil} 
\author{R. Menezes}\email{rmenezes@dce.ufpb.br}\affiliation{Departamento de Ci\^encias Exatas, Universidade Federal da Para\'\i ba, 58297-000 Rio Tinto, PB, Brazil}\affiliation{Departamento de F\'\i sica, Universidade Federal da Para\'\i ba, 58051-970 Jo\~ao Pessoa PB, Brazil}
\author{I. Zafalan}\email{ivzafalan@gmail.com}\affiliation{Departamento de F\'\i sica, Universidade Federal da Para\'\i ba, 58051-970 Jo\~ao Pessoa PB, Brazil} 

\begin{abstract}
{This work develops a procedure to find classes of Lagrangian densities that describe generalizations of the Abelian Maxwell-Higgs, the Chern-Simons-Higgs and the Maxwell-Chern-Simons-Higgs models. The investigation focuses on the construction of models that support vortices that obey the stressless condition and lead to first order differential equations which are compatible with the equations of motion. The results induce the appearance of constraints that restrict the choice of the Lagrangian densities, but help us to introduce an auxiliary function that allows to calculate the energy without knowing the explicit form of the solutions.}\\
\end{abstract}
\maketitle

\section{Introduction}

Topological structures appear in physics in several different contexts. They are spatially localized solutions with finite energy which attain topological properties that depend crucially on the number of spatial dimensions of the system in consideration. In high energy physics, the best known topological structures are kinks, vortices and monopoles, which appear in one, two and three spatial dimensions, respectively. In the standard scenario, kinks require a single real scalar field, vortices need a complex scalar field coupled to an Abelian gauge field, and monopoles appear in the presence of non Abelian gauge fields. For more details on this see, e.g., Refs.~\cite{b1,b2,b3} and references therein.

In the current work, we concentrate on vortices in relativistic models described in $(2,1)$ spacetime dimensions. As it is known, vortices were firstly studied by Helmholtz in Ref.~\cite{helmholtz} and are commonly found in fluid mechanics \cite{fluidmec}. They are also present in condensed matter when one studies superconductors. As it is well-known, when superconductors are below a critical temperature they expel the magnetic field, a phenomenom known as the Meissner effect \cite{meissner}. However, working with the Ginzburg-Landau theory of superconductivity \cite{glvortex}, in Ref.~\cite{abrikosov} Abrikosov noticed that vortices also appear in type II superconductors when exposed to an external electromagnetic field in a specific range of values.

The Ginzburg-Landau theory is nonrelativistic. Nevertheless, it is possible to find relativistic field theories that support vortex solutions. The first model was proposed by Nielsen and Olesen in Ref.~\cite{novortex}; it consists of a complex scalar field minimally coupled with a Maxwell gauge field under under the action of the local $U(1)$ gauge symmetry. The equations of the fields that describe the problem, however, are of second order involving the two aforementioned fields that interact in a nontrivial way. Even so, by setting an additional condition to the stress tensor, a classical solution was found in Ref.~\cite{vega} by de Vega and Schaposnik. Moreover, in Ref.~\cite{bogo} Bogomol'nyi found first order differential equations compatible with the equations of motion by minimizing the energy of the field configurations. The procedure developed by Bogomol'nyi works for kinks, vortices and monopoles, and joined the work of Prasad and Sommerfield \cite{ps} on monopoles, to make what is now called BPS states, which represent solutions of first order differential equations that solve the equations of motion and minimize the energy of the non-trivial static field configurations that describe the topological structures.  

Nielsen-Olesen vortices do not present electric charge. However, vortices with non vanishing electric charge were simultaneously found by Hong, Kim, and Pac in Ref.~\cite{coreanos} and by Jackiw and Weinberg in Ref.~\cite{csjackiw}. This interesting possibility of describing electrically charged vortex solutions appeared when instead of using the Maxwell term, one takes the Chern-Simons term for the dynamics of the gauge field in the Lagrangian density, which presents many exclusive properties in $(2,1)$ spacetime dimensions, as one can see in Refs.~\cite{csprop1,csprop2}. Chern-Simons vortices also support a first order formalism which minimizes the energy, as one can see in Ref.~\cite{csjackiw}.

Vortices including both Maxwell and Chern-Simons terms were firstly considered in Ref.~\cite{paulkhare} and later in \cite{intmcs}, where the interaction of two of them where studied. The equations of motion associated to these vortices are of second order. Here, however, it is not possible to reduce them to first order equations. Nevertheless, it was shown in Ref.~\cite{nmcs} that an additional neutral scalar field is needed in order to get first order equations that minimizes the energy and are compatible with the equations of motion. Other studies about this issue were done later in Refs.~\cite{bazeiamcs,susymcs}.

Over the years, non canonical models have also been considered to generate topological structures. The motivation arises in particular in the context of inflation in Ref.~\cite{ginf}. Later, they were proposed as solution for the cosmic coincidence problem \cite{ccoinc1,ccoinc2}. The features of these generalized models are in general different from the ones that appear for the standard models. For instance, there may be no need of a potential to drive the inflation of the universe in these models. In a context of topological defects, in Ref.~\cite{babichev1} Babichev studied the restrictions that models with global symmetry have to obey in order to support stable configurations. After that, several papers have dealt with similar issues, including kinklike solutions in flat spacetime \cite{gkink1,gkink2,gkink3} and in the braneworld scenario with a single extra dimension of an infinite extent \cite{gbrane1,gbrane2}. The formal studies of generalized gauged vortices started with Babichev in Ref.~\cite{bvortex}. Thenceforth, several papers have addressed this issue in various contexts \cite{dbibabichev,kvadam,clikebazeia,gcsbazeia,gbibazeia,gmhbazeia,gmcsbazeia,casana1,indonesios,casana2,atmaja}. However, as far as we know, the first paper that studied this issue is from $1991$ in Ref.~\cite{shiraishi}, where the authors found first order equations that minimizes the energy of the system to describe a Born-Infeld vortex. Some years later, Ref.~\cite{bimoreno} presented vortex solutions by adding the Chern-Simons term to the Lagrangian density.

More recently, vortices have been studied in several contexts, for instance, in massive gauged non-linear sigma models \cite{vv1}, as effective field theory for branes and strings carrying localized flux \cite{vv2}, in the study of non-Abelian vortices in holographic superconductors \cite{vv3}, in issues inspired by magnetic impurity considerations, in which some classes of Abelian Higgs and Chern-Simons-Higgs vortex equations are analyzed \cite{vv4}, in the search of vortices and magnetic bags in Abelian models with extended scalar sectors and some of their applications \cite{vv5}, as vortex configuration in the Abelian Higgs theory supplemented by higher order derivative self-interactions, related with Galileons \cite{vv6}, and also in terms of the volume of a vortex and the Bradlow bound \cite{vv7}.

In the present paper, we introduce a procedure to find general class of Lagrangian densities to describe generalizations of the Abelian Maxwell-Higgs (Sec.~\ref{maxwell}), the Chern-Simons-Higgs (Sec.~\ref{cs}) and the Maxwell-Chern-Simons-Higgs (Sec.~\ref{mcs}) vortices, that support stressless solutions which are compatible with the equations of motion. The study shows how to find specific constraints that restrict the choice of the Lagrangian density in the three distinct cases. We also show how to calculate the energy without knowing the explicit form of the solutions, and in Sec.~\ref{conclusions} we present our comments and conclusions.

\section{Maxwell-Higgs Vortices}
\label{maxwell}

We consider the generalized action for a gauge field and a complex scalar field in $(2,1)$ spacetime dimensions, with metric $\eta_{\mu\nu}={\rm diag}(+,-,-)$,
\be\label{gaction}
S=\int d^3x\LL(X,Y, |\vphi|),
\ee
where $\cal L$ is the Lagrangian density and
\be\label{XY}
X= \overline{D_\mu \vphi}D^\mu \varphi  \quad \text{and}\quad Y=-\frac14 F_{\mu\nu}F^{\mu\nu}.
\ee
Here, $\varphi$ stands for the complex scalar field and $A_\mu$ is the vector field. We also have $D_\mu = \partial_\mu +ieA_\mu$, $F_{\mu\nu}=\partial_\mu A_\nu - \partial_\nu A_\mu$ and the overline stands for the complex conjugation. The variation of the action \eqref{gaction} with respect to the fields $\vphi$ and $A_\mu$ gives the equations of motion
\bes\label{geom}
\begin{align}
 D_\mu (\LL_X D^\mu\vphi)&= \frac{\vphi}{2|\vphi|}\LL_{|\vphi|}, \\ \label{meqs}
 \partial_\mu \left(\LL_Y F^{\mu\nu} \right) &= J^\nu,
\end{align}
\ees
where the current is $J_\mu = ie{\cal L}_X(\bar{\vphi}D_\mu \vphi-\vphi\overline{D_\mu\vphi})$. We are also using the notation where ${\cal L}_X=\partial{\cal L}/\partial X,$ ${\cal L}_Y=\partial{\cal L}/\partial Y,$ and ${\cal L}_{|\varphi|}=\partial{\cal L}/\partial |\varphi|$. We can expand the above equations to get
\bes\label{eomexpmax}
\begin{align}
&\LX D_\mu D^\mu \vphi + 2\LL_{XX}D^\mu\vphi\,\Re\left(\overline{D_\alpha\vphi} \partial_\mu D^\alpha \vphi \right) -\frac12\LL_{XY}D^\mu\vphi F_{\alpha\beta}\partial_\mu F^{\alpha\beta}  \nonumber \\
 &= \frac{\vphi}{2|\vphi|}\LL_{|\vphi|} -\LL_{X|\vphi|}D^\mu\vphi\,\Re\left(\frac{\overline{\vphi}}{|\vphi|}\partial_\mu\vphi\right), \\
 &\LL_Y\partial_{\mu}F^{\mu\nu} + 2\LL_{YX}F^{\mu\nu}\,\Re\left(\overline{D_\alpha\vphi} \partial_\mu D^\alpha \vphi \right) -\frac12\LL_{YY}F^{\mu\nu} F_{\alpha\beta}\partial_\mu F^{\alpha\beta} \nonumber\\
 &= J^\nu -\LL_{Y|\vphi|}F^{\mu\nu} \,\Re\left(\frac{\overline{\vphi}}{|\vphi|}\partial_\mu\vphi\right).
\end{align}
\ees
In these equations we have used $\Re(z)$ to denote the real part of $z$. The energy-momentum tensor $T_{\mu\nu}$ for this generalized model is 
\be
T_{\mu\nu}=\LY F_{\mu\lambda}\tensor{F}{^\lambda_\nu}+ \LX\left( \overline{D_\mu \vphi}D_\nu \vphi + \overline{D_\nu \vphi}D_\mu \vphi\right) - \eta_{\mu\nu} \LL.
\ee
In the case of static solutions, the temporal component of Eq.~\eqref{meqs}, which is Gauss' law for our model, is solved with $A_0=0$. This makes the Maxwell-Higgs vortices electrically neutral \cite{novortex,vega}. Thus, the non-vanishing components of the energy-momentum tensor are given by
\bes\label{gTcomps}
\begin{align} \label{grho}
T_{00} &= -\LL, \\ 
T_{12} &= \LX \left(  \overline{D_1 \vphi}D_2 \vphi + \overline{D_2 \vphi}D_1 \vphi\right), \\
T_{11} &= \LY B^2 + 2\LX \left|D_1\vphi\right|^2 + \LL, \\ 
T_{22} &= \LY B^2 + 2\LX \left|D_2\vphi\right|^2 + \LL.
\end{align}
\ees
We take the usual ansatz for vortices
\bes\label{ansatz}
\begin{align}
\vphi(r,\theta)&=g(r)e^{i n\theta}, \\
A_i&=-\epsilon_{ij} \frac{x^j}{er^2}\left(a(r)-n\right),
\end{align}
\ees
where $r$ and $\theta$ are polar coordinates and $n$ is an integer number that represents the vorticity.  The functions $g(r)$ and $a(r)$ must obey the boundary conditions
\be\label{bcond}
g(0) = 0, \quad a(0)= n, \quad \lim_{r\to\infty} g(r) = v, \quad \lim_{r\to\infty} a(r) = 0,
\ee
where $v$ is the symmetry breaking parameter which has to be present in the model under investigation, such that the asymptotic values $a(\infty)=0$ and
$g(\infty)=v$ have to make the Lagrangian vanish and nullify the energy density. With the above ansatz, the functions $X$ and $Y$ in Eq.~\eqref{XY} become
\be\label{XYm}
X=-({g^\prime}^2+a^2g^2/r^2) \quad\text{and}\quad Y=-{a^\prime}^2/(2e^2r^2),
\ee
where the prime denotes the derivative with respect to $r$. Furthermore, the magnetic field assumes the form $B=-F^{12}=-a^\prime/(er)$ and the magnetic flux $\Phi=2\pi\int_0^\infty r dr B(r)$ is quantized:
\be\label{mflux}
\Phi=\frac{2\pi n}{e}.
\ee
The equations of motion \eqref{geom} with the ansatz \eqref{ansatz} become
\bes\label{secansatz}
\begin{align}\label{secansatzg}
\frac{1}{r} \left(r\LX g^\prime\right)^\prime -\frac{\LX a^2g}{r^2} + \frac12 \LL_{|\vphi|} &= 0, \\\label{secansatza}
r\left(\LY\frac{a^\prime}{r} \right)^\prime - 2e^2\LX ag^2 &= 0.
\end{align}
\ees
We can expand them as it was done in Eqs.~\eqref{eomexpmax} to get
\bes\label{secexpansatz}
\begin{align}\label{secexpansatzg}
 \LX g^{\prime\prime} +  \left( \LXX X^\prime + \LL_{XY} Y^\prime + \LL_{X|\vphi|}g^\prime +\frac{\LX}{r}\right)g^\prime -\frac{\LX a^2g}{r^2} + \frac12 \LL_{|\vphi|} &= 0, \\\label{secexpansatza}
\LY a^{\prime\prime} +\left(\LL_{YX}X^\prime + \LL_{YY}Y^\prime + \LL_{Y|\vphi|} g^\prime -\frac{\LL_Y}{r}\right)a^\prime - 2e^2\LX ag^2 &= 0.
\end{align}
\ees
In the above equations, we have $X^\prime = -(2g^\prime g^{\prime\prime} + 2a^2gg^\prime/r^2 + 2aa^\prime g^2/r^2 - 2a^2g^2/r^3)$ and $Y^\prime = -(2a^\prime a^{\prime\prime}/r^2 - 2{a^\prime}^2/r^3)/(2e^2)$. Regarding the energy-momentum tensor, the component \eqref{grho}, which is the energy density, does not change its explicit form with the ansatz \eqref{ansatz}. However, the other components of Eqs.~\eqref{gTcomps} take the forms
\bes
\begin{align}
T_{12} &= \LX \left( {g^\prime}^2 - \frac{a^2g^2}{r^2} \right) \sin(2\theta), \\ 
T_{11} &= \LY \frac{{a^\prime}^2}{e^2r^2} + 2\LX \left({g^\prime}^2\cos^2\theta+\frac{a^2g^2}{r^2}\sin^2\theta \right) + \LL, \\ 
T_{22} &= \LY \frac{{a^\prime}^2}{e^2r^2} + 2\LX \left({g^\prime}^2\sin^2\theta+\frac{a^2g^2}{r^2}\cos^2\theta \right) + \LL.
\end{align}
\ees
From the above equations, we can see that, in general, $T_{11} \neq T_{22}$. By using the rotation matrix in polar coordinates, it is possible to show that $T_{r\theta}=0$, $T_{rr}=\LY{a^\prime}^2/(er)^2 + 2\LX {g^\prime}^2 + \LL$ and $T_{\theta\theta}=\LY{a^\prime}^2/(er)^2 + 2\LX a^2g^2/r^2 + \LL$. Then, in the polar coordinate system, the spatial components of the energy momentum tensor do not depend on the angle $\theta$. 

We perform the reescale $r\to z=\lambda r$ in the solutions $a(r)$ and $g(r)$, which makes $X \to X^{(\lambda)} = \lambda^2 X(z)$ and $Y \to Y^{(\lambda)} = \lambda^4 Y(z)$, where $X(z)$ and $Y(z)$ mean to change $r\to z$ in Eq.~\eqref{XYm}. The energy of the reescaled functions can be calculated; it is given by
\be
\begin{split}
E^{(\lambda)} &= -2\pi \int_0^\infty r dr \LL\left(X^{(\lambda)}, Y^{(\lambda)}, g^{(\lambda)}\right)\\
&= -\frac{2\pi}{\lambda^2} \int_0^\infty z dz \LL\left(\lambda^2 X(z), \lambda^4 Y(z), g(z)\right).
\end{split}
\ee
The above energy must have a minimum at $\lambda=1$ because it has to be minimized for the non-reescaled solution. Then, the condition $\left.\partial E^{(\lambda)}/\partial \lambda\right|_{\lambda=1} =0$ lead us to
\be\label{intderrick}
\int_0^\infty z dz \left(\LL - \LX X - 2\LY Y\right) = 0.
\ee
The above equation is the condition that makes the solutions stable under reescaling. We look back to the energy momentum tensor components and suppose that $T_{\mu\nu}$ is axially symmetric, i.e., angle-independent, by taking
\be\label{godeq}
{g^\prime}^2 =  \frac{a^2g^2}{r^2}.
\ee
This makes $X=-2{g^\prime}^2 = -2a^2g^2/r^2$. By taking this into account, we get $T_{12}= T_{21}=0$ and
\be
T_{11} = T_{22} = \LY \frac{{a^\prime}^2}{e^2r^2} + 2\LX {g^\prime}^2 + \LL.
\ee
Since the energy-momentum tensor is conserved, we have $T_{11}=T_{22}=C$, where $C$ is a constant. By using Eq.~\eqref{intderrick}, we see that $T_{11}=T_{22}=0$. Then, we get the equation
\be\label{derrickcond}
\LL - \LX X - 2\LY Y = 0.
\ee
Notice that this approach is different from the one considered in Ref.~\cite{vega}, which uses the stressless condition, $T_{ij}=0$, as an ansatz. Here, we suppose that the energy momentum tensor is axially symmetric and show that the stressless condition is a consequence of the stability under rescaling.

The condition in Eq.~\eqref{derrickcond} constrains the functions $a$, $g$, $a^\prime$ and $g^\prime$. One can derivate the above expression to get
\be\label{dderrick}
2g^\prime\left(\frac{1}{r} \left(r\LX g^\prime\right)^\prime -\frac{\LX a^2g}{r^2} + \frac12 \LL_{|\vphi|} \right) +  \frac{a^\prime}{e^2r^2}\left(r\left(\LY\frac{a^\prime}{r} \right)^\prime - 2e^2\LX ag^2\right) = 0,
\ee
which contains the equations of motion \eqref{secansatz}. Combining Eq.~\eqref{godeq} with Eq.~\eqref{derrickcond}, we get a first order differential equation for $a$, in the form of an algebraic equation, $M(a^\prime/r,a/r,g)= 0$. We can study the behavior of the solutions near the origin by taking 
\be\label{orim}
a_0(r)\approx n-\tilde{a}_0(r) \quad\text{and}\quad g_0(r)\approx \tilde{g}_0(r),
\ee
by considering terms up to first order in $\tilde{g}_0(r)$ and $\tilde{a}_0(r)$. Combining this with Eq.~\eqref{godeq}, we get that $g_0\propto r^n$, regardless the specific form of the Lagrangian density. However, the behavior of $a_0(r)$ is not general, since it is obtained by Eq.~\eqref{derrickcond}, which explicitly depends on the model in consideration.

We then use the first order equations \eqref{godeq} and \eqref{derrickcond} in the energy density \eqref{grho} to get
\be\label{rhofom}
\rho = \LY \frac{{a^\prime}^2}{e^2r^2} + 2\LX {g^\prime}^2.
\ee
From the above equation, we see that the energy density depends only on the derivatives of the functions $a(r)$ and $g(r)$. This fact is important since it allows that we introduce an auxiliary function $W=W(a,g)$ and consider
\be\label{wawg}
W_a = \LY \frac{a^\prime}{e^2r} \quad \text{and} \quad W_g = 2\LX r g^\prime,
\ee
where $W$ must obey the constraint $W_g=2\LX ag$ to be compatible with Eq.~\eqref{godeq}. In this case, the energy density can be written as
\be
\rho=\frac{1}{r} \frac{dW}{dr},
\ee
and the energy is then given by
\be\label{energywmax}
E = 2\pi \left|W\left(a(\infty),g(\infty)\right)-W\left(a(0),g(0)\right)\right|.
\ee
The boundary conditions that appear from Eq.~{\eqref{bcond} shows that $E=2\pi|W(0,v)-W(n,0)|$. This is an important novelty of the procedure, and shows that the first order formalism not only makes the job of getting solutions easier, but also allows to calculate the energy exactly, without even knowing the solutions. The issue here is then how to calculate $W=W(a,g)$, but the answer depends on the specific model under consideration, as we further comment below.

To use the stated first order equations, one needs to check their compatibility with the equations of motion \eqref{secansatz}. In particular, by taking the square root of \eqref{godeq} and considering $g^\prime = ag/r$, Eq.~\eqref{secexpansatzg} becomes
\be\label{constraintmL}
\frac{a^\prime}{r}g\LX + \frac{ag}{r}\left(\LXX X^\prime + \LL_{XY} Y^\prime + \LL_{X|\vphi|}\frac{ag}{r}\right) + \frac12 \LL_{|\vphi|}=0,
\ee
where $X^\prime$ and $Y^\prime$ denotes the radial derivative of the functions $X$ and $Y$ as in Eq.~\eqref{XYm}. By using the first order equation \eqref{godeq}, we get $X^\prime = -4(a^2g^2(a-1)/r^3 + aa^\prime g^2/r^2)$. Since we have used the equation of motion \eqref{secansatzg} to find the above constraint, one could infer that the equation of motion \eqref{secansatza} lead to another constraint. However, the above equation implies that, if it is satisfied and the first order equations \eqref{godeq} and \eqref{derrickcond} are used, we see from Eq.~\eqref{dderrick} that the equation of motion \eqref{secansatza} is satisfied.

The Eq.~\eqref{constraintmL} shows that not every Lagrangian density supports stressless solutions, only the ones that satisfy it. It gives an equation in the form $I\left(a^{\prime\prime}, a^\prime, a, g, r\right)=0$. However, if it is possible, one can use the first order equation \eqref{derrickcond} to isolate $a^\prime$ as a function of $a$, $g$ and $r$, so that Eq.~\eqref{constraintmL} becomes a constraint in the form $I\left(a, g, r\right)=0$. In general, this constraint can only be tested after the solutions were found from the first order equations \eqref{godeq} and \eqref{derrickcond}, by taking $I\left(a(r), g(r), r\right)=0$. A way of knowing that the solutions satisfy the constraint \textit{a priori} is to interpret it as a differential equation. This can be done by considering Lagrangian densities that lead to a constraint that is a differential equation whose variable is $g$. This happens if the constraint $I\left(a, g, r\right)=0$ depends only on $g$. In this case, it is possible to calculate how the scalar field must appear in the Lagrangian density. However, the term with $\LXX$ and $\LL_{XY}$ makes the construction of analytical models very hard because it is not possible to explicitly write $r$ and $a$ in terms of $g$. For instance, considering the Lagrangian density
\be
\LL = -\alpha X^2 + Y - V(|\vphi|),
\ee
with $\alpha$ being a real constant, we see that Eq.~\eqref{constraintmL} becomes
\be\label{exruimm}
\frac{12\alpha a^2g^3}{r^3} a^\prime + \frac{8\alpha a^3g^3}{r^4}\left(a-1\right) = \frac12 V_{|\vphi|},
\ee
in which we have considered $g^\prime = ag/r$ from Eq.~\eqref{godeq}. An attempt to eliminate $r$ and $a$ in the above equation can be made by using Eq.~\eqref{derrickcond}, which gives
\be\label{dcondruim}
\frac{{a^\prime}^2}{2e^2r^2} = V-\frac{4\alpha a^4g^4}{r^4}.
\ee
We then take $a^\prime = -er\sqrt{2V-8\alpha a^4g^4/r^4}$ in Eq.~\eqref{exruimm} to get
\be
-\frac{12\alpha ea^2g^3}{r^2} \sqrt{2V-\frac{8\alpha a^4g^4}{r^4}} + \frac{8\alpha a^3g^3}{r^4}\left(a-1\right) = \frac12 V_{g}.
\ee
Therefore, although one can solve the problem using the first order equations \eqref{godeq} and \eqref{dcondruim}, it is impossible to find the potential analytically, since $r$ and $a$ appear in the above constraint. Of course one can use a numerical approach to calculate the potential from the above equation, in a way that the Lagrangian density that satisfies the above constraint is only numerical, without an analytical expression for the potential. Notice that, if the solution is known, one may write $r$ and $a$ as functions of $g$ and determine the potential in the above equation. However, the purpose of this paper is to find the most general class of Lagrangian densities that allows to be constructed without knowing the analytical stressless solutions. We then see that the only possibility to do so is by taking linear expressions on $X$, with a factor that depends on the scalar field. We then work with the Lagrangian density
\be\label{lgenpossible}
\LL = K(|\vphi|) X + G(Y,|\vphi|). 
\ee 
In this case, the constraint \eqref{constraintmL} becomes:
\be\label{constraintmG}
\frac{a^\prime}{r}gK + \frac12 G_{|\vphi|}=0.
\ee
We can use Eq.~\eqref{derrickcond} to get the equation
\be\label{derrickdm}
G-2YG_Y=0.
\ee
This is an algebraic equation for $Y$ as a function of $|\vphi|$, and here we recall the recent investigation \cite{VVV}, which deals with similar issues. Therefore, we can use the explicit form of $Y$ in Eq.~\eqref{XYm} and take
\be\label{eqeffm}
-Y=\frac{{a^\prime}^2}{2e^2r^2} = V_{eff}(|\vphi|),
\ee
where $V_{eff}(|\vphi|)$ is an effective potential associated to the theory. By using the above equation and taking $a^\prime/r = -e\sqrt{2V_{eff}}$ into Eq.~\eqref{constraintmG}, we get the constraint between the functions:
\be\label{bestconstm}
2egK\sqrt{2V_{eff}} -G_{|\vphi|} = 0.
\ee
Here, $G_{|\vphi|}$ is just a function of $g$ because its implicit dependence on $Y$ is now written in terms of the effective potential as in Eq.~\eqref{eqeffm}. This equation can be used to construct the Lagrangian density. Note, however, that we only have used Eq.~\eqref{secansatzg}. In this case, one can show that Eqs.~\eqref{godeq}, \eqref{derrickdm}, \eqref{eqeffm} and the constraint \eqref{bestconstm} automatically solve the equation of motion \eqref{secansatza}, as expected from Eq.~\eqref{dderrick}. For Lagrangian densities written in the form \eqref{lgenpossible}, we get that Eqs.~\eqref{wawg}, combined with Eqs.~\eqref{godeq} and \eqref{eqeffm} admit the function
\be\label{wagm}
W(a,g) = -\frac{a}{e}\left(G_Y\sqrt{-2Y}\right)_{Y=-V_{eff}}
\ee
if the constraint \eqref{bestconstm} is satisfied. Thus, if the effective potential is known, the energy can be calculated analytically. The result shows that for models of the form \eqref{lgenpossible}, the function $W$ is given by the above expression.

To exemplify, we take the standard model
\be\label{lstandard}
K(|\vphi|) =1 \quad\text{and}\quad G(Y,|\vphi|) = Y - V(|\vphi|).
\ee
To find the effective potential that appears in the constraint \eqref{bestconstm}, we must solve Eq.~\eqref{derrickdm} for $-Y$ to get $V_{eff}(|\vphi|)=V(|\vphi|)$. By using Eq.~\eqref{bestconstm}, one can show that the only potential that is allowed in the standard case is the potential 
\be\label{phi4}
V(|\vphi|) = \frac{e^2}{2}(v^2-|\vphi|^2)^2.
\ee
With this potential, the first order equations are
\be\label{fostdm}
{g^\prime} = \frac{ag}{r} \quad\text{and}\quad {a^\prime} = -e^2r(v^2-g^2).
\ee
In this case, the energy density in Eq.~\eqref{grho} is given by
\be
\begin{split}
\rho &= -X - Y + V \\
     &=\frac{2a^2g^2}{e^2r^2} + e^2(v^2-g^2)^2.
\end{split}
\ee
This is the first relativistic model considered for Maxwell vortices. Its solutions were studied in Refs.~\cite{novortex,vega,bogo} by numerical analysis and have energy $E=2\pi|n|v^2$. 

Considering the class of models \eqref{lgenpossible}, it is possible to generalize the standard model with a general function $K(|\vphi|)$ and
\be\label{gen1}
G(Y,|\vphi|) = -P(|\vphi|)(-Y)^s-V(|\vphi|),
\ee
where $s$ is a real parameter such that $s>1/2$. The standard case is easily obtained for $P(|\vphi|)=K(|\vphi|)=1$, and $s=1$. In the newer case, the effective potential obtained by solving Eq.~\eqref{derrickdm} for $-Y$ is
\be
V_{eff}(|\vphi|) = \left(\frac{V(|\vphi|)}{(2s-1)P(|\vphi|)}\right)^{1/s}.
\ee
The constraint \eqref{bestconstm} written in terms of the above effective potential becomes
\be\label{const1m}
\left(2V_{eff}\frac{dP}{dg} +(2s-1)P\frac{dV_{eff}}{dg}\right)sV_{eff}^{s-1} = -2egK\sqrt{2V_{eff}},
\ee
which leads to the potential
\be\label{potpkm}
V(|\vphi|) = (2s-1)P(|\vphi|)\left(\frac{ev^2 - e\int_0^{|\vphi|}d\tilde{g}\, 2\,\tilde{g}\,K(\tilde{g})}{s\sqrt{2}P(|\vphi|)}\right)^{\frac{2s}{2s-1}},
\ee
where $v^2$ is the parameter involved in the symmetry breaking. The above equation shows that not any function $K(|\vphi|)$ is possible to be chosen because it has to allow the symmetry breaking of the potential. We then impose that $V(v)=0$. We want to point out that the above class of potentials is not the only solution possible for the constraint in Eq.~\eqref{const1m}. We have considered it here because for $K(|\vphi|)=P(|\vphi|)=s=1$ one naturally recovers the standard case illustrated in Eq.~\eqref{phi4}. In this case, the first order equations are 
\be\label{eqspkm}
{g^\prime} =  \frac{ag}{r} \quad\text{and}\quad a^\prime = -er\left(\frac{ev^2 - e\int_0^{g}d\tilde{g}\, 2\,\tilde{g}\,K(\tilde{g})}{s\,2^{1-s}P(g)}\right)^{\frac{1}{2s-1}}.
\ee

To illustrate this example we consider the case in which
\be\label{illustm}
s=2/3,\quad P(|\vphi|)=\alpha,\quad K(|\vphi|)=l\,(|\vphi|/v)^{2l-2},
\ee
where $\alpha$ is a constant that keeps the dimension of the term $P(|\vphi|) (-Y)^s$ fixed, and $l$ is a real parameter such that $l\geq1$. In this case, the potential \eqref{potpkm} becomes
\be\label{potexm}
V(|\vphi|) = \frac{27e^4v^8}{64\alpha^3}\left(1-\left(\frac{|\vphi|}{v}\right)^{2l}\right)^4.
\ee
The above potential reminds the one studied in Ref.~\cite{fktc}, which was used to compactify kinklike solutions. The route proposed there led to compact vortices in Ref.~\cite{compvortex}. However, in the work \cite{compvortex} the compactification happens because of the function $P(|\vphi|)$, in a manner that the function $K(|\vphi|)$ plays no role. In this paper, we show how the function $K(|\vphi|)$ works to change the behavior of the vortex with the parameter $l$, leaving $P(|\vphi|)$ unchanged. The potential \eqref{potexm} has its minima located at $|\vphi|=v$ and its local maximum is located at $|\vphi| = 0$, such that $V_{\max}=V(0) = 27e^4v^8/64\alpha^3$. This behavior does not change, regardless the $l$ that is chosen. In Fig.~\ref{figvm}, we plot the above potential for $e,v,\alpha=1$, and $l=1$ and increasing to larger values. One can see that a plateau appears between $0\leq|\vphi|\leq v$ and becomes wider as $l$ increases.
\begin{figure}[!htb]
\centering
\includegraphics[width=6cm]{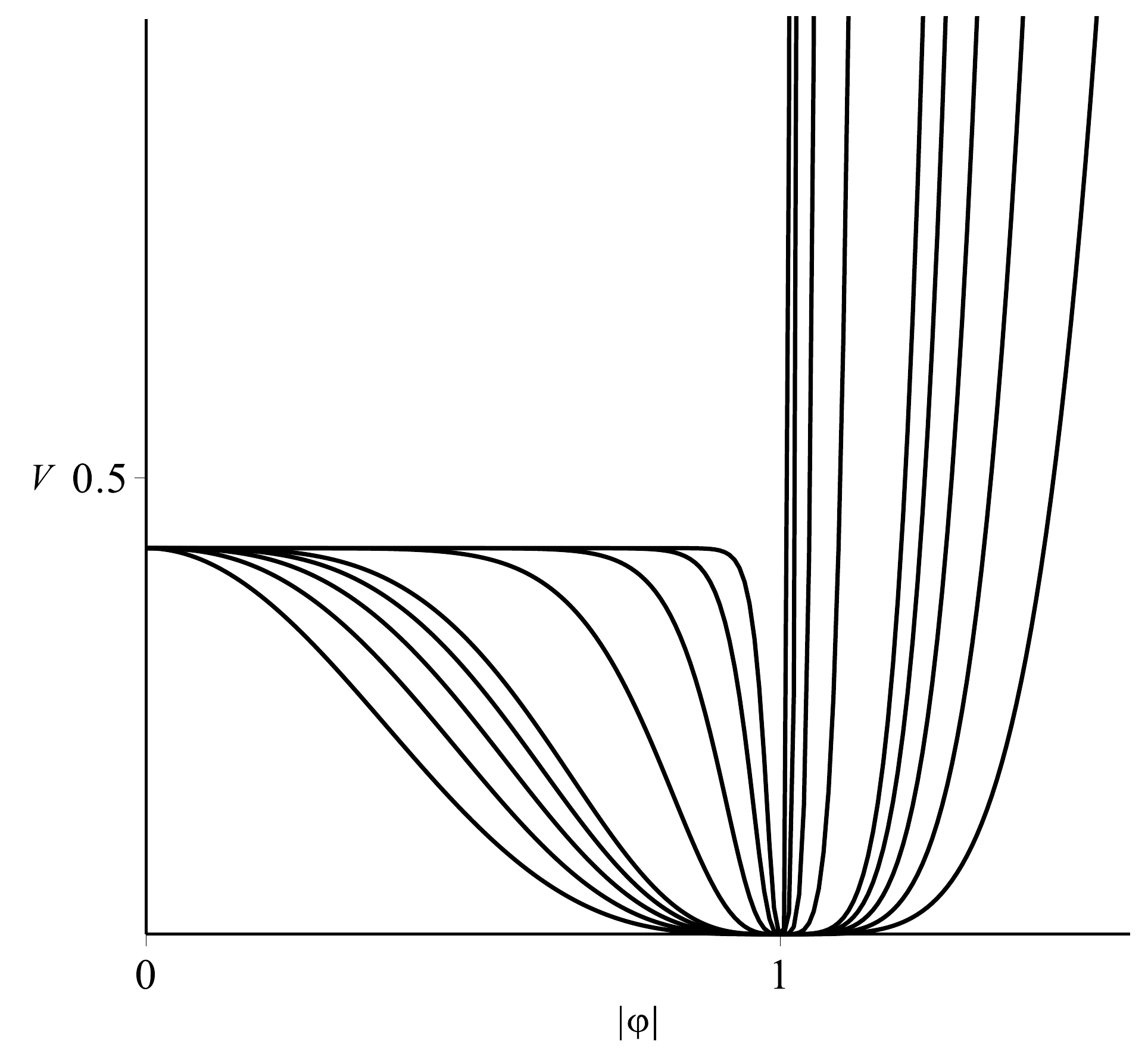}
\caption{The potential of Eq.~\eqref{potexm} plotted for $e,v,\alpha=1$, and $l=1,2,4,8,16$ and $32$. The plateau becomes wider as $l$ increases.}
\label{figvm}
\end{figure} 

In this case, the first order equations \eqref{eqspkm} become
\be\label{foexm}
{g^\prime} =  \frac{ag}{r} \quad\text{and}\quad a^\prime = -\frac{27e^4v^6}{16\alpha^3}\,r\left(1-\left(\frac{g}{v}\right)^{2l}\right)^3.
\ee
Before going further, we investigate the behavior of the solutions near the origin by using Eq.~\eqref{orim} to get
\be\label{oriexm}
a_0(r)\approx n-\frac{27e^4v^6}{32\alpha^3}r^2 \quad\text{and}\quad g_0(r)\approx C r^n,
\ee
where $C$ is a real constant to be ajusted to the curve. Finding analytical solutions for this set of first order equations \eqref{foexm} is not an easy task. So, we have conducted our investigation mainly numerically. Nevertheless, using the above first order equations we have been able to find analytical expressions for $l\to\infty$, given by
\bes\label{solcexm}
\bal
{a}_\infty(r)&=
\begin{cases}
n-\frac{27e^4v^6}{32\alpha^3}r^2,\,\,\,& r\leq r_c\\
0, \,\,\, & r>r_c.
\end{cases} \\
{g}_\infty(r)&=
\begin{cases}
v\left(\frac{r}{r_c}\right)^n \exp\left(\frac{27e^4v^6}{64\alpha^3}(r_c^2-r^2) \right) ,\,\,\, & r\leq r_c\\
v, \,\,\, & r>r_c.
\end{cases}
\eal
\ees
In the above expressions, $r_c = 4\sqrt{6\alpha^3 n}/(9e^2v^3)$ is the compactification radius, since both
${a}_\infty(r)$ and ${g}_\infty(r)$ reach the asymptotic values at finite $r$. However, they are not differentiable, so they cannot be solutions of the equations of motion. In Fig.~\ref{solspkm}, however, we have plotted the numerical solutions for $e,v,\alpha,n=1$, and several values of $l$, including both ${a}_\infty(r)$ and ${g}_\infty(r)$ to illustrate how the solutions behave asymptotically.
\begin{figure}[!htb]
\centering
\includegraphics[width=6cm]{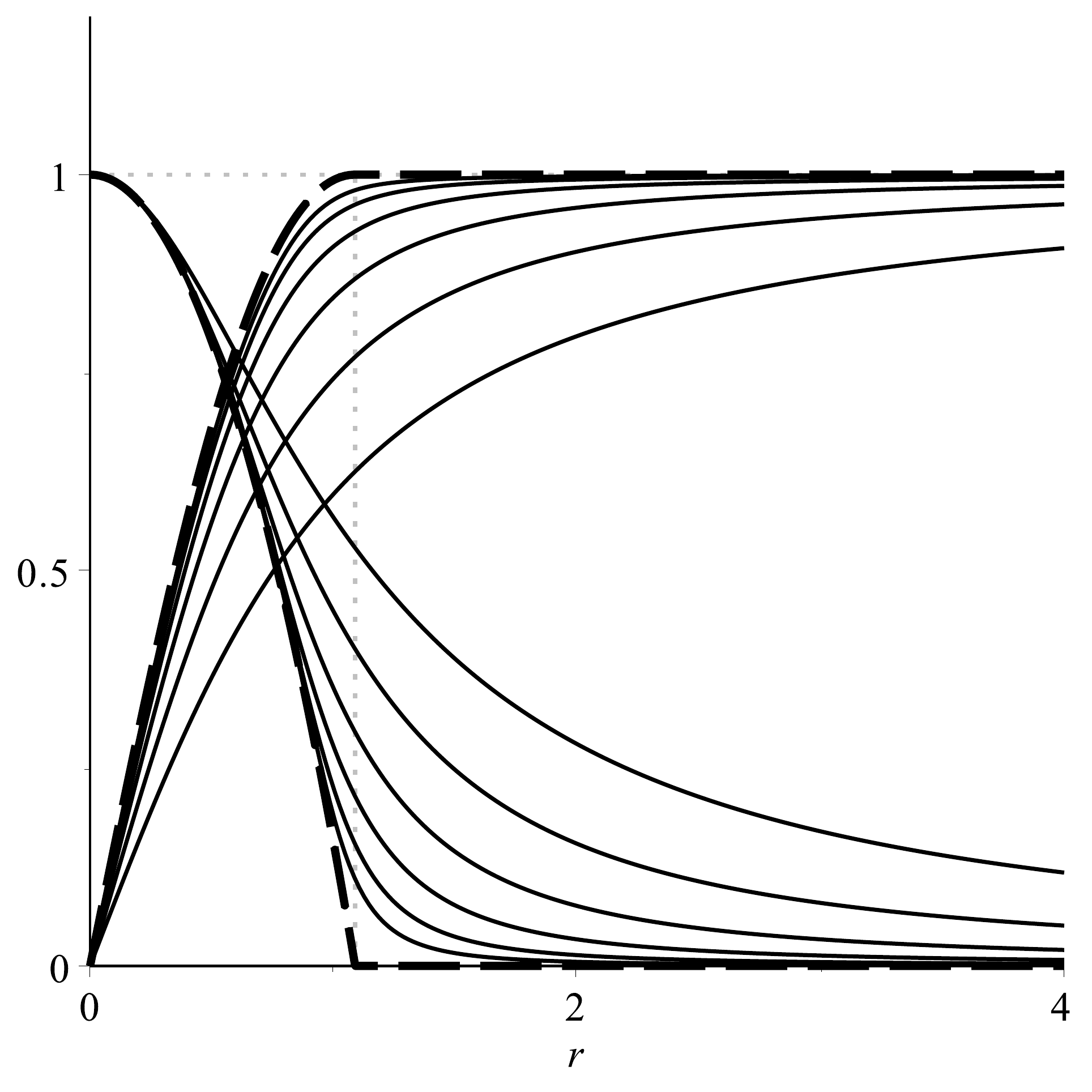}
\caption{The solutions $a(r)$ and $g(r)$ of Eqs.~\eqref{foexm} plotted for $e,v,\alpha,n=1$,  and $l=1,2,4,8,16,32$ and the expressions in Eqs.~\eqref{solcexm} represented by the dashed lines. The ascending lines stand for the function $g(r)$, whilst the other ones are for $a(r)$.}
\label{solspkm}
\end{figure} 

The magnetic field, given by $B(r)=-a^\prime/(er)$ is also calculated numerically, for a general $l$.  In Fig.~\ref{bspkm} we show the magnetic field for $e,v,\alpha,n=1$, and several values of $l$, including the limit $l\to \infty$.
\begin{figure}[!htb]
\centering
\includegraphics[width=6cm]{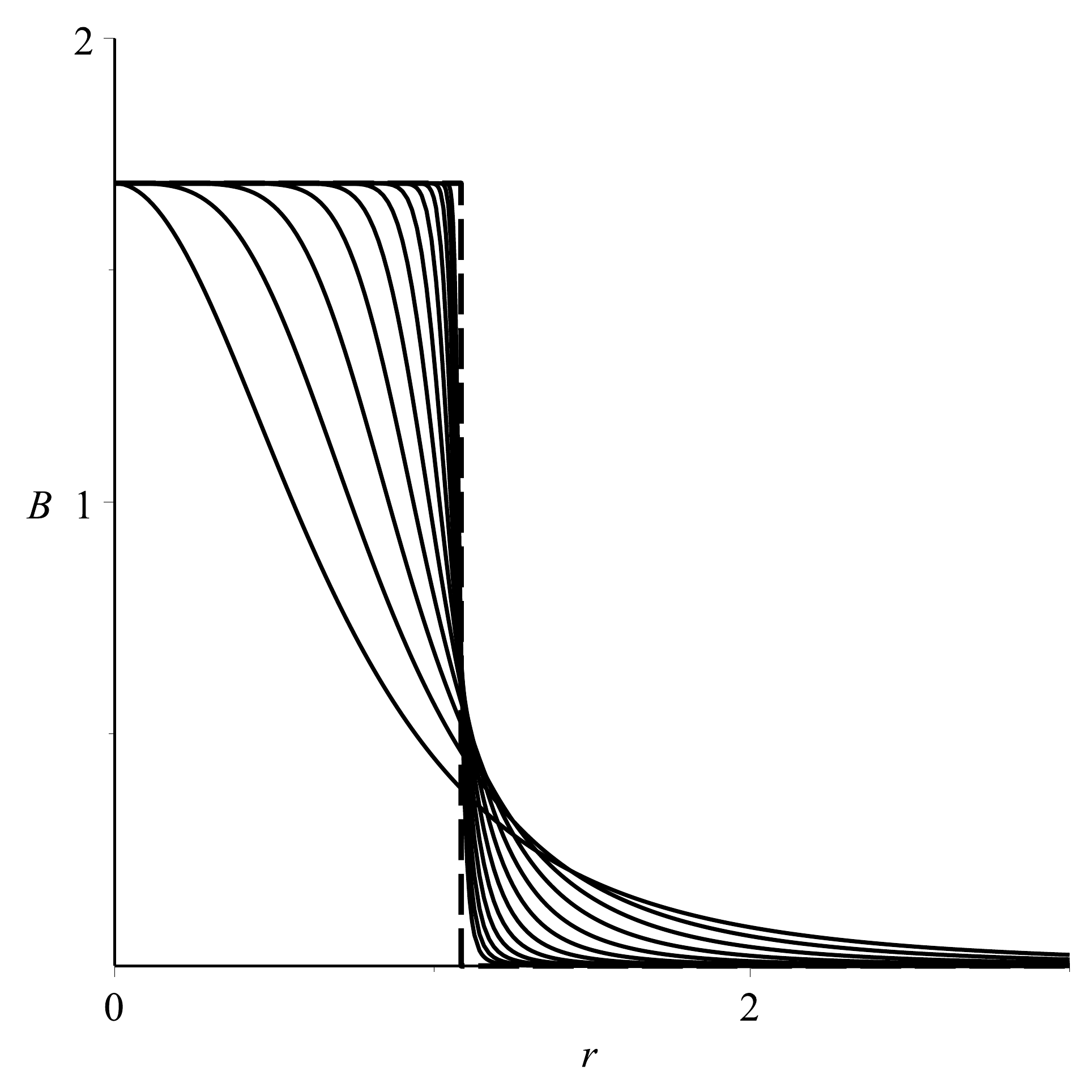}
\caption{The magnetic field $B(r)=-a^\prime/(er)$ associated to the solutions of Eqs.~\eqref{foexm} plotted for $e,v,\alpha,n=1$,  and $l=1,2,4,8,16,32,64,128,256,512,2048$ and the asymptotic limit, $l\to\infty$, represented by the dashed line.}
\label{bspkm}
\end{figure} 

To investigate the energy density, we use Eq.~\eqref{grho} to get
\be\label{rhoexm}
\rho(r) = \alpha\left(\frac{{a^\prime}}{\sqrt{2}er}\right)^{4/3} + 2l\left(\frac{g}{v}\right)^{2l-2} {g^\prime}^2 + \frac{27e^4v^8}{64\alpha^3}\left(1-\left(\frac{g}{v}\right)^{2l}\right)^4.
\ee
The above equations must be combined with the numerical solutions of the first order equations \eqref{foexm}. In Fig.~\ref{rhospkm} we have plotted the energy density for $e,v,\alpha,n=1$, and several values of $l$, including the asymptotic limit $l\to\infty$. One can see that the energy density tends to become uniform inside the compact space $0\leq r\leq r_c$ as the parameter $l$ gets larger and larger. The inset shows its behavior near the origin for $1\leq l\leq2$, which is non perturbative with $l$. This can be checked analytically with the help of the expressions for the functions $a(r)$ and $g(r)$ near the origin as in Eqs.~\eqref{oriexm} with $C=\beta v$, which lead to 
\be
\rho^{r\approx0}_l(r) = \frac{81e^4v^8}{64\alpha^3} + 2ln^2v^2\beta^{2l}r^{2nl-2} + \frac{27e^4v^8}{64\alpha^3}\left(1-\beta^{2l} r^{2nl}\right)^4.
\ee
The above expression shows that, for $n=1$ and $l=1$, $\rho^{r\approx0}_1(0)=27e^4v^8/(16\alpha^3)+2\beta^2$, and for $n=1$ and $l\neq1$, $\rho^{r\approx0}_l(0)=27e^4v^8/(16\alpha^3)$. We then see that, near the origin the behavior of the solution as a function of $l$ is nonperturbative.
\begin{figure}[t]
\centering
\includegraphics[width=6cm]{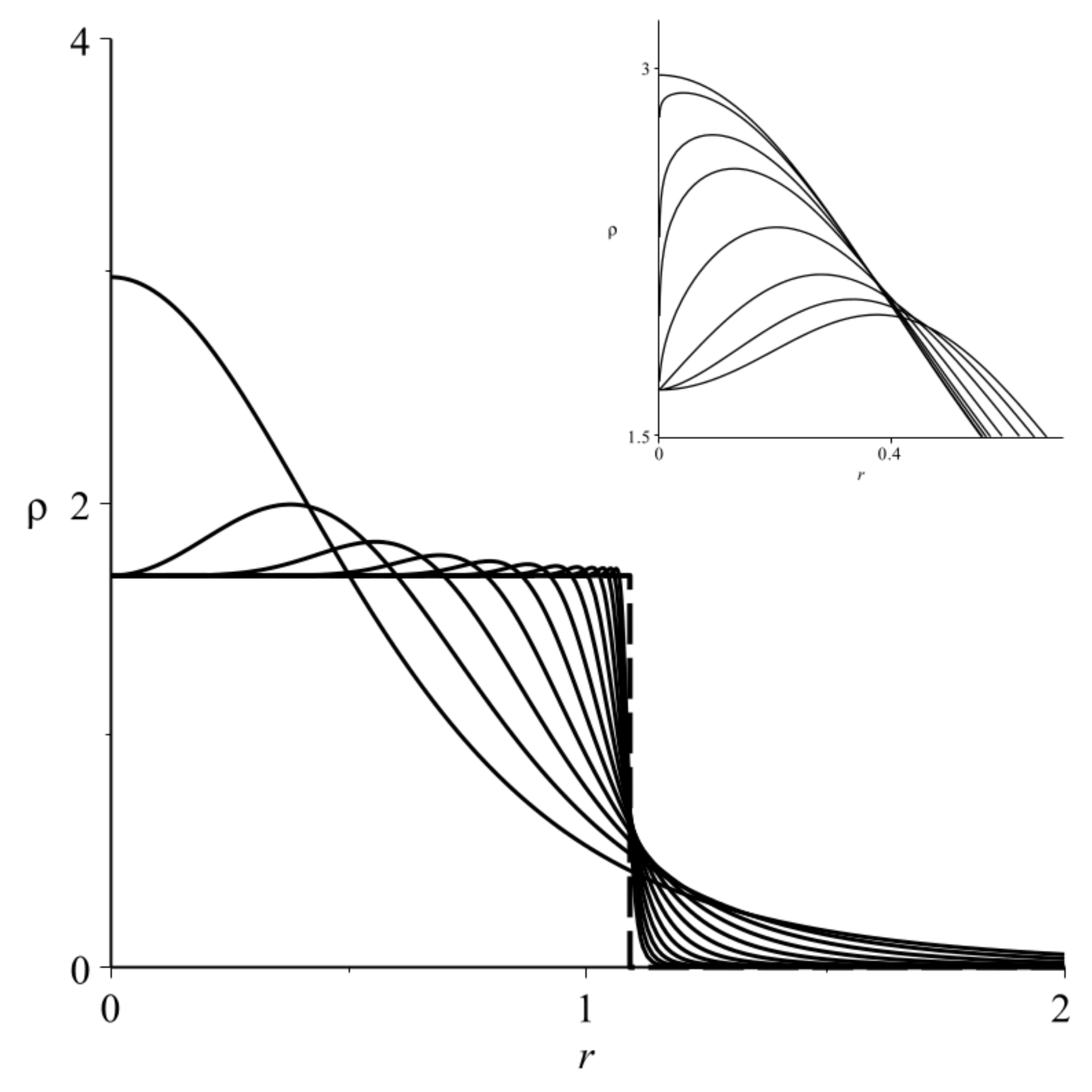}
\caption{The energy density \eqref{rhoexm} associated to the solutions of Eqs.~\eqref{foexm} plotted for $e,v,\alpha,n=1$, and $l=1,2,4,8,16,32,64,128,256,512,2048$ and the asymptotic limit $l\to\infty$, represented by the dashed line. The inset shows the behavior of the energy density near the origin for $1\leq l\leq2$.}
\label{rhospkm}
\end{figure} 
It is worth commenting that, without using the formalism \eqref{energywmax}, the energy can be calculated through numerical integration to get $E\approx2\pi$ for  $e,v,\alpha,n=1$. At the end of this section we show that this energy calculated by direct integration can be easily obtained with the use of the auxiliar function $W(a,g)$ defined by Eq.~\eqref{wagm}. 

In order to highlight the behavior of the vortex configuration described above, in Fig.~\ref{fig5} we depict the energy density in the
$(r,\theta)$ plane for some values of $l$, including the limit of a very large value of $l$. We see from the results in Fig.~\ref{fig5}, that the effect of increasing $l$ that makes the vortex shrink to a compact disklike region works very smoothly, and this is different from the case of the Chern-Simons vortex configuration which we study in the next section.

Another possibility is to study a Born-Infeld model, inspired in Ref.~\cite{shiraishi}, which reminds the ALTW model for kinks \cite{trodden}; see also \cite{VVV}. To do so, we take $K(|\vphi|)=1$ and
\be\label{gbialtw}
G(Y,|\vphi|) = M^2-M^2\sqrt{\left(1+\frac{2V(|\vphi|)}{M^2}\right)\left(1-\frac{2Y}{M^2}\right)}.
\ee
In this case, by using Eq.~\eqref{derrickdm} we get the effective potential
\be
V_{eff}(|\vphi|) = V(|\vphi|).
\ee
and the constraint \eqref{bestconstm} becomes the same of the standard case, which gives the potential \eqref{phi4} and first order equations \eqref{fostdm}. One can use Eq.~\eqref{grho} to show that the energy density is given by
\be
\begin{split}
\rho &= -X +M^2\sqrt{\left(1+\frac{2V(|\vphi|)}{M^2}\right)\left(1-\frac{2Y}{M^2}\right)}-M^2 \\
     &=\frac{2a^2g^2}{e^2r^2} + e^2(v^2-g^2)^2.
\end{split}
\ee
Therefore, since this model and the standard model \eqref{lstandard} admit the same stressless solutions and energy densities, they are twins.
\begin{figure}[t!]
\centering
\includegraphics[width=3.5cm]{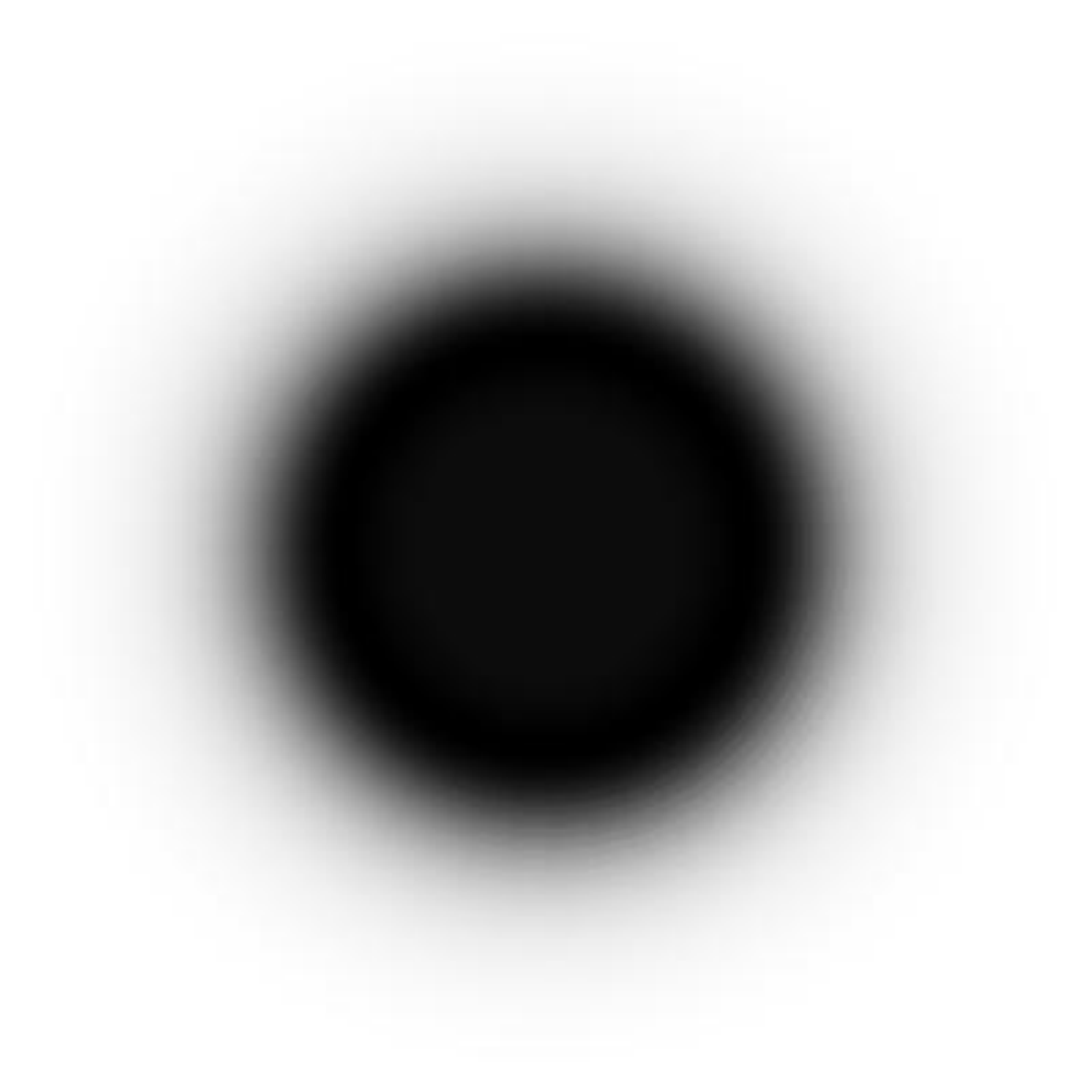}
\includegraphics[width=3.5cm]{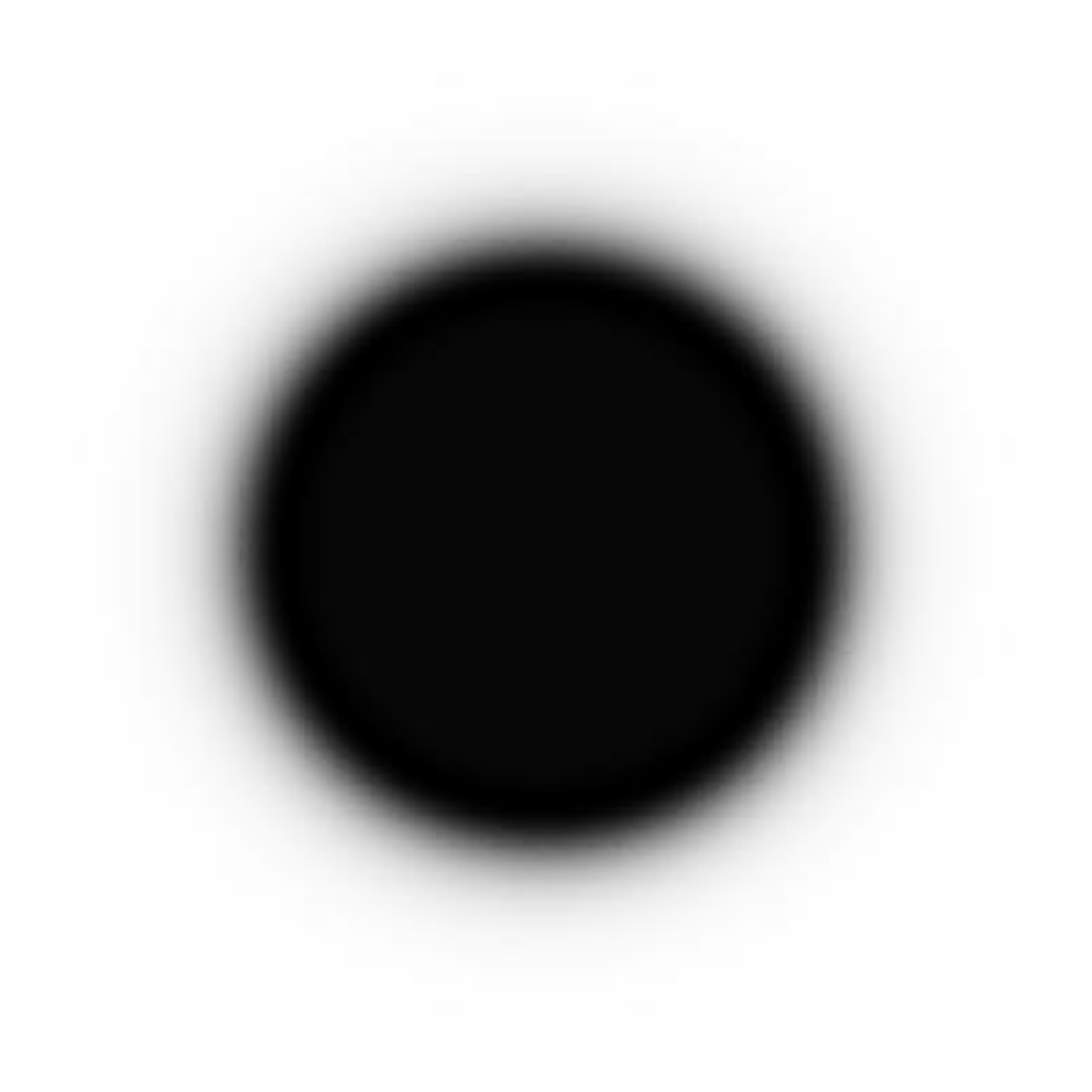}
\includegraphics[width=3.5cm]{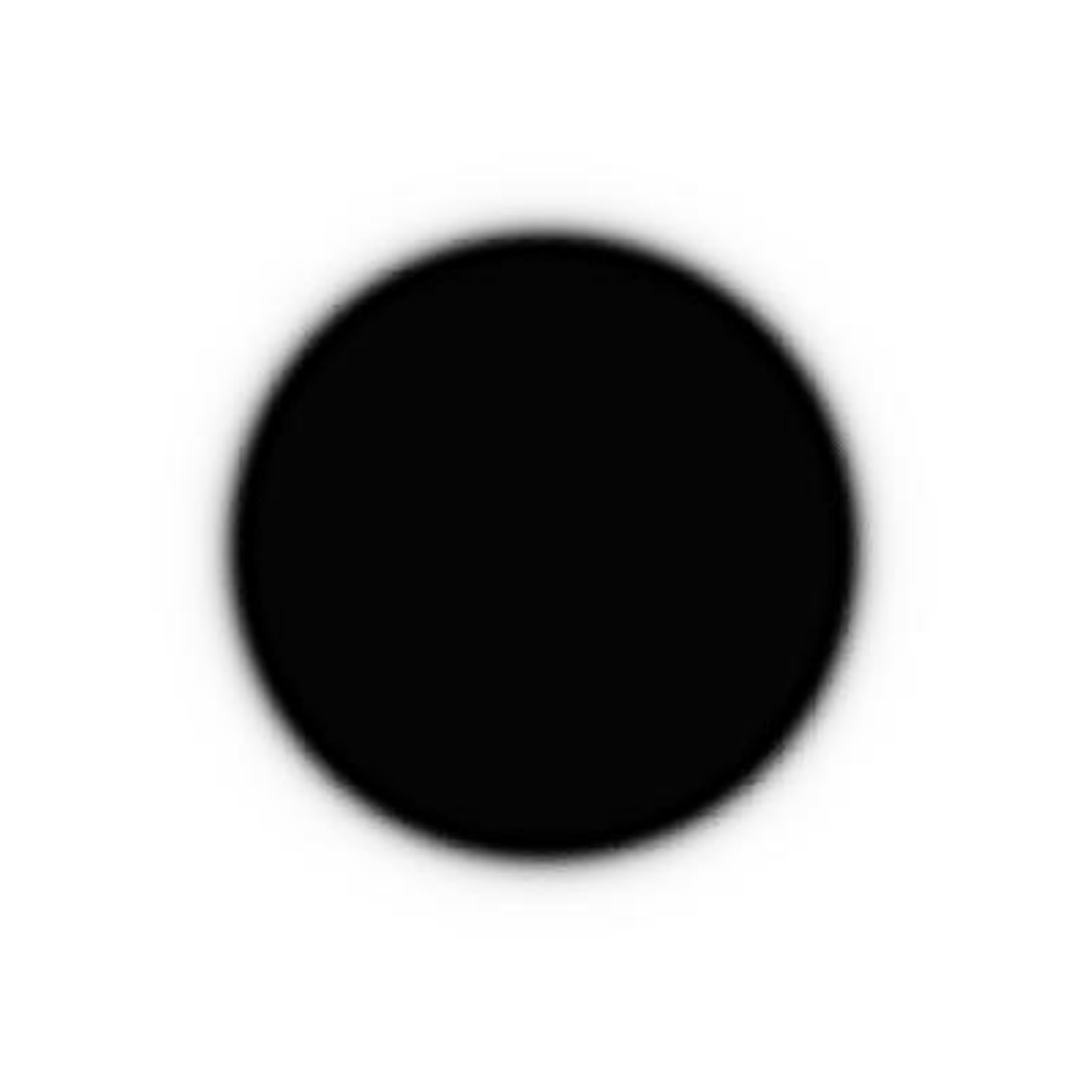}
\includegraphics[width=3.5cm]{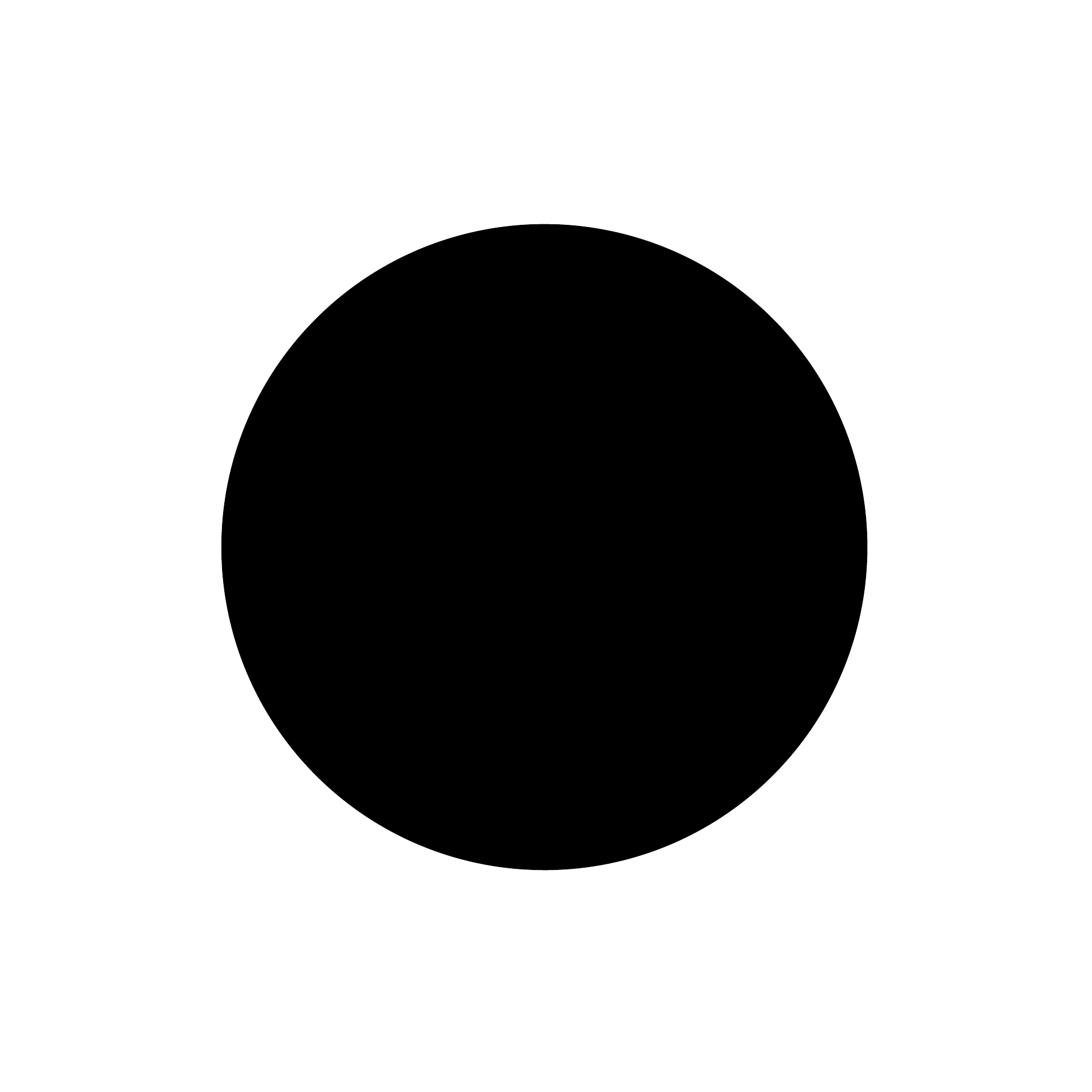}
\caption{The energy density \eqref{rhoexm} associated to the solutions of Eqs.~\eqref{foexm} displayed in the $(r,\theta)$ plane for $e,v,\alpha,n=1$, and $l=8,32,256$ and for a much larger value of $l$.}
\label{fig5}
\end{figure} 

We now return to the issue of calculation of the function $W$, turning attention to the energy of the models presented above, which are all in the class \eqref{lgenpossible}. We have shown in Eq.~\eqref{energywmax} that it is possible to use the function $W=W(a,g)$ to calculate the energy without knowing the explicit form of the solution. By using Eq.~\eqref{wagm}, we have that the function $W$ is given by
\be 
 W(a,g)=a\left(\int_0^{g} d\tilde{g}\, 2\tilde{g}\,K(\tilde{g}) -v^2\right),
\ee
for the model \eqref{gen1}. In the model \eqref{illustm} it is
\be
 W=- a v^2 + a v^2 \left(\frac{g}{v}\right)^{2l}.\ee
Also, it is
\be 
W=a(g^2-v^2),
\ee
for the model \eqref{gbialtw}. In both cases, considering that $W(0,v)=0$, the energy becomes
$E=2\pi |n| v^2$ and attains the same value that appears in the standard model \eqref{lstandard}. This is a consequence of the form of the potential that we have chosen in Eq.~\eqref{potpkm}. The energy may be different for other potentials that solve the constraint in Eq.~\eqref{const1m}.

\section{Chern-Simons-Higgs Vortices}\label{cs}
We now consider the most general action for the Chern-Simons system
\be\label{gcsaction}
S=\int d^3x\left[\LL(X, |\vphi|) + \frac{\kappa}{4}\epsilon^{\alpha\beta\gamma}A_\alpha F_{\beta\gamma}\right],
\ee
where $X$ is as in Eq.~\eqref{XY}. We cannot consider a generalization of the Chern-Simons term in the Lagrangian density because it would break gauge invariance. The variation of the action \eqref{gcsaction} with respect to the fields $\vphi$ and $A_\mu$ gives the equations of motion
\bes\label{gcseom}
\begin{align}
 D_\mu (\LL_X D^\mu\vphi)&= \frac{\vphi}{2|\vphi|}\LL_{|\vphi|}, \\ \label{cseqs}
 \frac{\kappa}{2} \epsilon^{\lambda\mu\nu}F_{\mu\nu} &= J^\lambda,
\end{align}
\ees
where the current is $J_\mu = ie{\cal L}_X(\bar{\vphi}D_\mu \vphi-\vphi\overline{D_\mu\vphi})$. It is possible to expand the equation of the scalar field to get
\be\label{gcseomexp}
\LX D_\mu D^\mu \vphi + 2\LL_{XX}D^\mu\vphi\,\Re\left(\overline{D_\alpha\vphi} \partial_\mu D^\alpha \vphi \right) = \frac{\vphi}{2|\vphi|}\LL_{|\vphi|} -\LL_{X|\vphi|}D^\mu\vphi\,\Re\left(\frac{\overline{\vphi}}{|\vphi|}\partial_\mu\vphi\right).
\ee
The energy momentum tensor $T_{\mu\nu}$ for the generalized model \eqref{gcsaction} is given by 
\be
T_{\mu\nu}=\LX\left( \overline{D_\mu \vphi}D_\nu \vphi + \overline{D_\nu \vphi}D_\mu \vphi\right) - \eta_{\mu\nu} \LL.
\ee
The components of the energy momentum tensor for static solutions are given by
\bes\label{gTcscompss}
\begin{align}\label{gcsrho}
T_{00} &= 2\LX e^2 A_0^2 |\vphi|^2-\LL, \\ 
T_{0i} &= -A_0 J_i, \\
T_{12} &= \LX \left(  \overline{D_1 \vphi}D_2 \vphi + \overline{D_2 \vphi}D_1 \vphi\right), \\
T_{11} &= 2\LX \left|D_1\vphi\right|^2 + \LL, \\ 
T_{22} &= 2\LX \left|D_2\vphi\right|^2 + \LL,
\end{align}
\ees
In the case of static solutions, differently from the Maxwell vortices of the previous case, we cannot solve the Gauss' law - the temporal component of Eq.~\eqref{cseqs} - with $A_0=0$ anymore, so the Chern-Simons vortices are electrically charged. This implies that $T_{0i}\neq0$: because the Chern-Simons vortices are electrically charged, they produce a current density that carries momentum. We take the ansatz \eqref{ansatz} with the boundary conditions \eqref{bcond} and consider $A_0$ a radial function, which makes the electric field given by $(E_x,E_y)\equiv E^i=\partial^i A_0 =-{A_0}^\prime x^i/r$. In this case, $X=e^2g^2A_0^2-({g^\prime}^2+a^2g^2/r^2)$ and, since we have not changed the ansatz, the magnetic field still is $B=-F^{12}=-a^\prime/(er)$ and the magnetic flux also obeys the quantization in Eq.~\eqref{mflux}. From the temporal component of the Eq.~\eqref{cseqs}, we get that $A_0$ is constrained to obey
\be\label{A0}
A_0 = \frac{\kappa}{2\LX e^2} \frac{B}{|\vphi|^2}.
\ee
This makes the Chern-Simons vortices charged \cite{coreanos,csjackiw}. The electric charge is given by $Q =2\pi\int rdr J^0$. The above equation allows us to show that the charge can be written in terms of the magnetic flux \eqref{mflux} as $Q = -\kappa \Phi$, which makes the electric charge quantized. The equations of motion \eqref{gcseom} with the ansatz \eqref{ansatz} and $A_0=A_0(r)$, are given by
\bes\label{eomcsansatz}
\begin{align}
\frac{1}{r} \left(r\LX g^\prime\right)^\prime + \LX g \left(e^2 A_0^2-\frac{a^2}{r^2} \right) + \frac12 \LL_{|\vphi|} &= 0, \\ \label{a0csansatz}
 \frac{a^\prime}{r} + \frac{2\LX e^3g^2 A_0}{\kappa} &= 0, \\ \label{a0pcs}
 {A_0^\prime} + \frac{2\LX ea g^2}{\kappa r} &= 0.
\end{align}
\ees
The above equation of motion for the scalar field can be expanded as in Eq.~\eqref{gcseomexp}, that leads to
\be\label{eomcsexpansatz}
\LX g^{\prime\prime} +  \left( \LXX X^\prime + \LL_{X|\vphi|}g^\prime +\frac{\LX}{r}\right)g^\prime + \LX g \left(e^2 A_0^2-\frac{a^2}{r^2} \right) + \frac12 \LL_{|\vphi|} = 0.
\ee
As before, the component \eqref{gcsrho} of the energy momentum tensor, which is the energy density, does not change its explicit form with the ansatz \eqref{ansatz}. However, the other components of Eqs.~\eqref{gTcscompss} becomes
\bes\label{gTcscomp}
\begin{align}
T_{01} &= -\frac{2\LX e ag^2 A_0 \sin{\theta}}{r},  \\
T_{02} &= \frac{2\LX e ag^2 A_0 \cos{\theta}}{r}, \\
T_{12} &= \LX \left( {g^\prime}^2 - \frac{a^2g^2}{r^2} \right) \sin(2\theta), \\ 
T_{11} &= 2\LX \left({g^\prime}^2\cos^2\theta+\frac{a^2g^2}{r^2}\sin^2\theta \right) + \LL, \\
T_{22} &= 2\LX \left({g^\prime}^2\sin^2\theta+\frac{a^2g^2}{r^2}\cos^2\theta \right) + \LL.
\end{align}
\ees
We now take stressless solutions as done in the previous section. This leads us to Eq.~\eqref{godeq}, which comes from $T_{12}=0$. We can perform a similar reescaling in the energy as the one done in the last section to show that $T_{11}=T_{22}=0$, which lead us to
\be\label{derrickcs}
\LL+2\LX {g^\prime}^2 = 0.
\ee
One can derivate the above equation with respect to $r$ and use the first order equation \eqref{godeq} and the Gauss' Law \eqref{a0csansatz}, to get
\be\label{dderrickcs}
2g^\prime\left( \frac{1}{r} \left(r\LX g^\prime\right)^\prime + \LX g \left(e^2 A_0^2-\frac{a^2}{r^2} \right) + \frac12 \LL_{|\vphi|} \right) - \frac{\kappa a^\prime}{er} \left( {A_0^\prime} + \frac{2\LX ea g^2}{\kappa r} \right) = 0.
\ee
The above equation presents the same fact of Eq.~\eqref{dderrick} for Maxwell-Higgs vortices: it contains the equations of motion \eqref{eomcsansatz}, except the Gauss' Law \eqref{a0csansatz}, which was used with \eqref{godeq} to find it. By considering the first order equations \eqref{derrickcs} and \eqref{a0csansatz}, we can write the energy density \eqref{gcsrho} as
\be
\rho = \frac{\kappa^2}{2\LX e^4g^2} \frac{{a^\prime}^2}{r^2} + 2\LX {g^\prime}^2.
\ee
Then, as in the Maxwell vortices, we see that the energy density depends only on the derivatives of the functions $a(r)$ and $g(r)$. We can use an auxiliary function $W=W(a,g)$ and take
\be\label{wgcs}
W_a = \frac{\kappa^2}{2\LX e^4g^2} \frac{{a^\prime}}{r} \quad \text{and} \quad W_g = 2\LX r g^\prime.
\ee
However, $W$ must obey the constraint $W_g=2\LX ag$ to be compatible with Eq.~\eqref{godeq}. Moreover, Eq.~\eqref{a0csansatz} allows us to write $W_a=-\kappa A_0/e$. In this case, the energy density can be written as
\be
\rho=\frac{1}{r} \frac{dW}{dr},
\ee
and the energy is given by
\be\label{energyw}
E = 2\pi \left|W\left(a(\infty),g(\infty)\right)-W\left(a(0),g(0)\right)\right|.
\ee
This possibility is similar to the case studied before in the previous section, so we omit further details here.

The stressless condition provides the first order equations Eqs.~\eqref{godeq} and \eqref{derrickcs}. However, we need to find solutions for three functions: $A_0(r),a(r),g(r)$. Thus, we must use Eqs.~\eqref{godeq} and \eqref{derrickcs} with one of the equations of motion \eqref{eomcsansatz} to solve the problem. We can see then that two of the equations of motion may constrain these functions. Before going further, we have to check the compatibility of the first order equations \eqref{godeq} and \eqref{derrickcs} with the equations of motion \eqref{eomcsansatz}. Using $g^\prime=ag/r$ in Eq.~\eqref{eomcsexpansatz} we get
\be\label{constraintcsV}
\frac{ag}{r} \LXX X^\prime + e^2A_0g\LX\left(A_0-
\frac{2e}{\kappa}g^2\LX\right) + \frac{a^2g^2}{r^2}\LL_{X|\vphi|} + \frac12\LL_{|\vphi|} = 0,
\ee
where $X^\prime$ stands for the radial derivative of the function $X$. By using Eq.~\eqref{dderrickcs}, one can show that the use of the first order equations \eqref{godeq}, \eqref{a0csansatz} and \eqref{derrickcs} in models that obey Eq.~\eqref{constraintcsV} implies that the equation of motion \eqref{a0pcs} is automatically solved. Furthermore, by using arguments similar to the previous section, we get from Eq.~\eqref{constraintcsV} that a path to construct the Lagrangian density analytically is to consider models with linear dependencies in $X$ in the Lagrangian density. Then, the most general Lagrangian density that allows to be constructed analytically is
\be\label{lgenkcs}
\LL = K(|\vphi|)X - V(|\vphi|),
\ee
where $K(|\vphi|)$ is an adimensional function. This model was studied in Ref.~\cite{gcsbazeia}. In this case, Eq.~\eqref{constraintcsV} becomes
\be\label{constra0cs}
e^2A_0^2gK-\frac{2e^3}{\kappa}g^3A_0K^2+\frac12 e^2A_0^2g^2K_g - \frac12 V_g=0.
\ee
From Eq.~\eqref{derrickcs}, we get the equation
\be\label{a0v}
e^2g^2A_0^2=\frac{V(g)}{K(g)}.
\ee
Plugging this into Eq. \eqref{constra0cs}, one can show that the potential and the function $K(|\vphi|)$ are related by the equation
\be\label{KVrel}
\frac{d}{dg}\left(\sqrt{\frac{V}{g^2K}}\right)=-\frac{2e^2}{\kappa}g K.
\ee
This equation is solved by the potential
\be\label{potgencs}
V(|\vphi|) = \frac{e^4}{\kappa^2}|\vphi|^2K(|\vphi|)\left(v^2-\int_0^{|\vphi|}d\tilde{g}\, 2\,\tilde{g}\,K(\tilde{g})\right)^2.
\ee
Again, as in the previous section, $v^2$ is the parameter involved in the symmetry breaking and not any function $K(|\vphi|)$ is possible because it has to allow the symmetry breaking of the potential. We then impose that $V(v)=0$. Also, as in the Maxwell-Higgs case, we have considered the above potential because it naturally recovers the standard case for $K(|\vphi|)=1$.

The first order equations to be solved in this case come from Eq.~\eqref{godeq} and the combination of Eqs.~\eqref{a0csansatz} and \eqref{a0v} for the potential \eqref{potgencs}
\be\label{foeqcs}
g^\prime=\frac{ag}{r} \quad\text{and}\quad a^\prime = -\frac{2e^4rg^2K(g)}{\kappa^2}\left(v^2-\int_0^{|\vphi|}d\tilde{g}\, 2\,\tilde{g}\,K(\tilde{g})\right).
\ee
By using the potential \eqref{potgencs} in Eqs.~\eqref{wgcs} with \eqref{a0v}, one can show that the function $W$ is given by
\be\label{wcs}
W(a,g)=-av^2+ a \int_0^{g}d\tilde{g}\, 2\,\tilde{g}\,K(\tilde{g}).
\ee
Thus, considering that $W(0,v)=0$, the energy for the model \eqref{lgenkcs} with potential \eqref{potgencs} is $E=2\pi|n|v^2$. This is a consequence of the form of the potential in Eq.~\eqref{potgencs}. Thus, other values of energy may be obtained for different potentials that solve the constraint \eqref{KVrel}.
\begin{figure}[t]
\centering
\includegraphics[width=6cm]{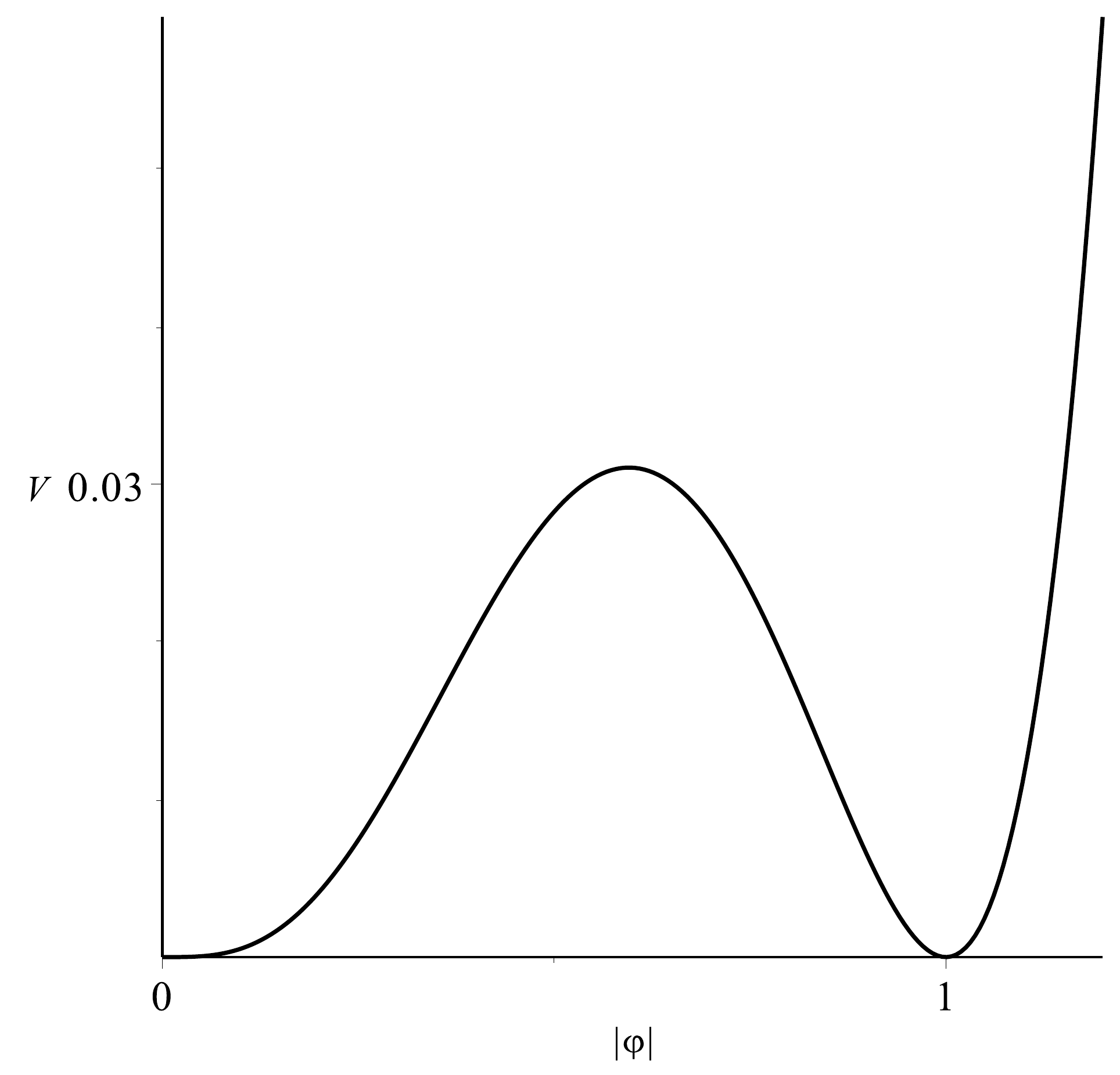}
\caption{The potential of Eq.~\eqref{vexe}, depicted for $e,v,\kappa=1$.}
\label{figvcs}
\end{figure} 

Some examples that falls into the class of models studied above can be found in Ref.~\cite{gcsbazeia}. In particular, we can include the new model, defined by
\bes
\bal
K(|\vphi|) &= \left(\frac{|\vphi|}{v}\right)^2\ln^2\left(\frac{|\vphi|}{v}\right), \\
V(|\vphi|) &= \frac{e^4v^6}{\kappa^2}\left(\frac{|\vphi|}{v}\right)^4 \ln^2\left(\frac{|\vphi|}{v}\right) \left(1-\frac{1}{16}\left(\frac{|\vphi|}{v}\right)^4 \left(8\ln^2\left(\frac{|\vphi|}{v}\right) - 4 \ln\left(\frac{|\vphi|}{v}\right) + 1\right)\right)^2\label{vexe}
\eal
\ees
These functions are compatible with the constraint \eqref{KVrel}. The above potential presents a minimum at $|\vphi|=0$ and another one at $|\vphi|=1$. In between these minima, there is a maximum which is hard to be calculated analytically. Nevertheless, for $e,v,\kappa=1$, we get that the maximum is located at $|\vphi|\approx 0.595$, and in Fig.~\ref{figvcs} we display this potential for $e,v,\kappa=1$. 

In this case, the first order equations \eqref{foeqcs} become
\be\label{focslog}
g^\prime=\frac{ag}{r} \quad\text{and}\quad a^\prime = -\frac{2e^4v^4r}{\kappa^2}\left(\frac{g}{v}\right)^4\ln^2\left(\frac{g}{v}\right)\left(1-\frac{1}{16}\left(\frac{g}{v}\right)^4 \left(8\ln^2\left(\frac{g}{v}\right) - 4 \ln\left(\frac{g}{v}\right) + 1\right)\right).
\ee
We solve them for $e,v,\kappa=1$ and show their behavior in Fig.~\ref{figsolcs}.
\begin{figure}[!htb]
\centering
\includegraphics[width=6cm]{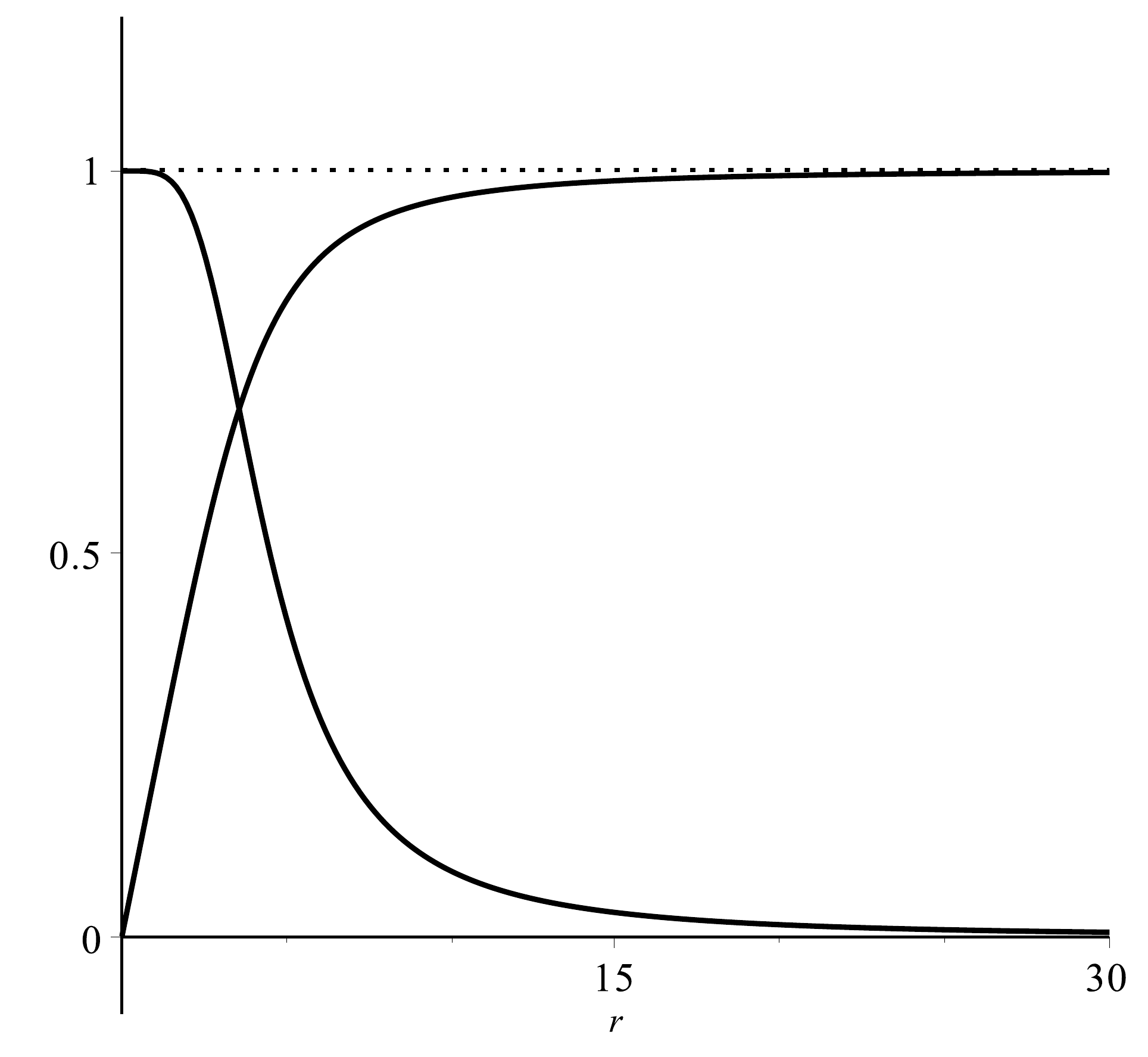}
\caption{The solutions $a(r)$ (descending line) and $g(r)$ (ascending line) of Eqs.~\eqref{focslog}, depicted for $e,v,\kappa=1$.}
\label{figsolcs}
\end{figure} 
The electric and magnetic fields are given by $E=A_0^\prime$ and $B=-a^\prime/(er)$. We depict them in Fig.~\ref{figebcs}
\begin{figure}[!htb]
\centering
\includegraphics[width=5cm]{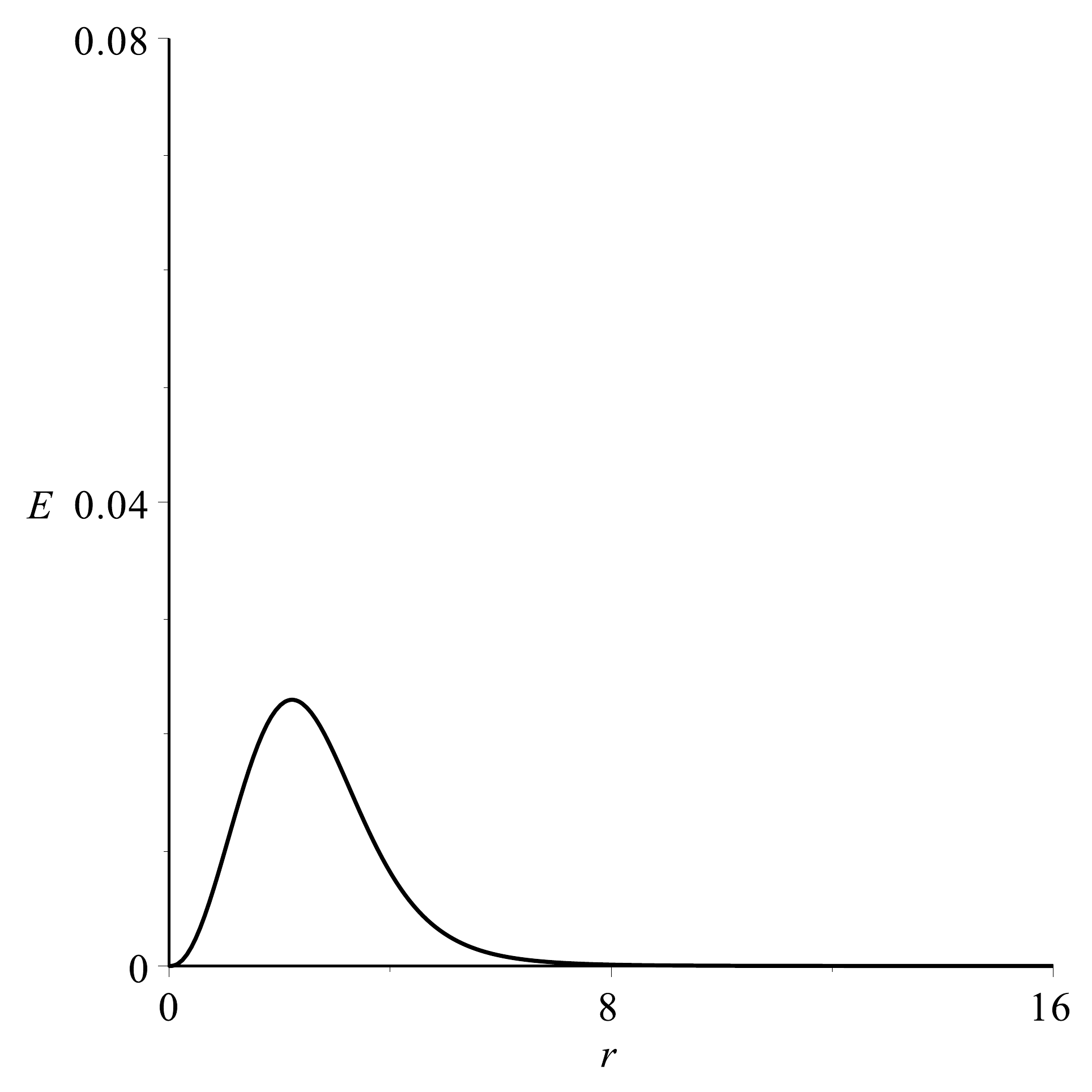}
\includegraphics[width=5cm]{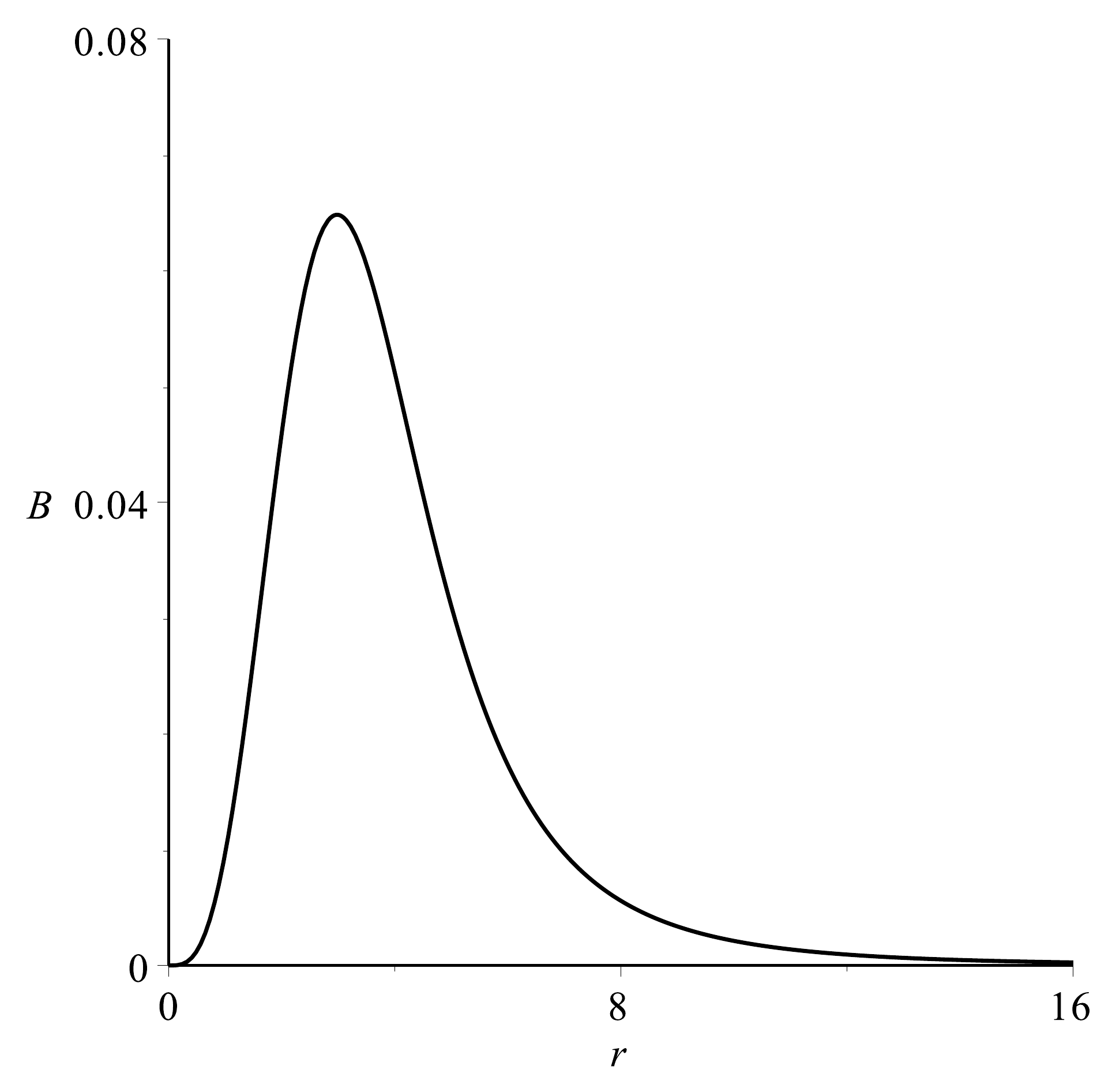}
\caption{The electric (left) and magnetic (right) fields for the solutions $a(r)$ and $g(r)$ of Eqs.~\eqref{focslog}, depicted for $e,v,\kappa=1$.}
\label{figebcs}
\end{figure} 

The energy density can be calculated to give
\be\label{rhologcs}
\rho = \frac{v^2\kappa^2}{2e^4g^4\ln^2\left(g/v\right) } \frac{{a^\prime}^2}{r^2} + 2\left(\frac{g}{v}\right)^2\ln^2\left(\frac{g}{v}\right) {g^\prime}^2.
\ee
In Fig.~\ref{figrhocs} we display the energy density for $e,v,\kappa=1$. The total energy can be calculated from $W=W(a,g)$, which in the current case is given by
\be 
W(a,g)=av^2\left(\frac1{16v^4}g^4\left(8\ln^2\left(\frac{g}{v}\right)-4\ln\left(\frac{g}{v}\right)+1 \right)-1 \right) 
\ee
It gives $E=2\pi |n|v^2.$ 
\begin{figure}[!htb]
\centering
\includegraphics[width=6cm]{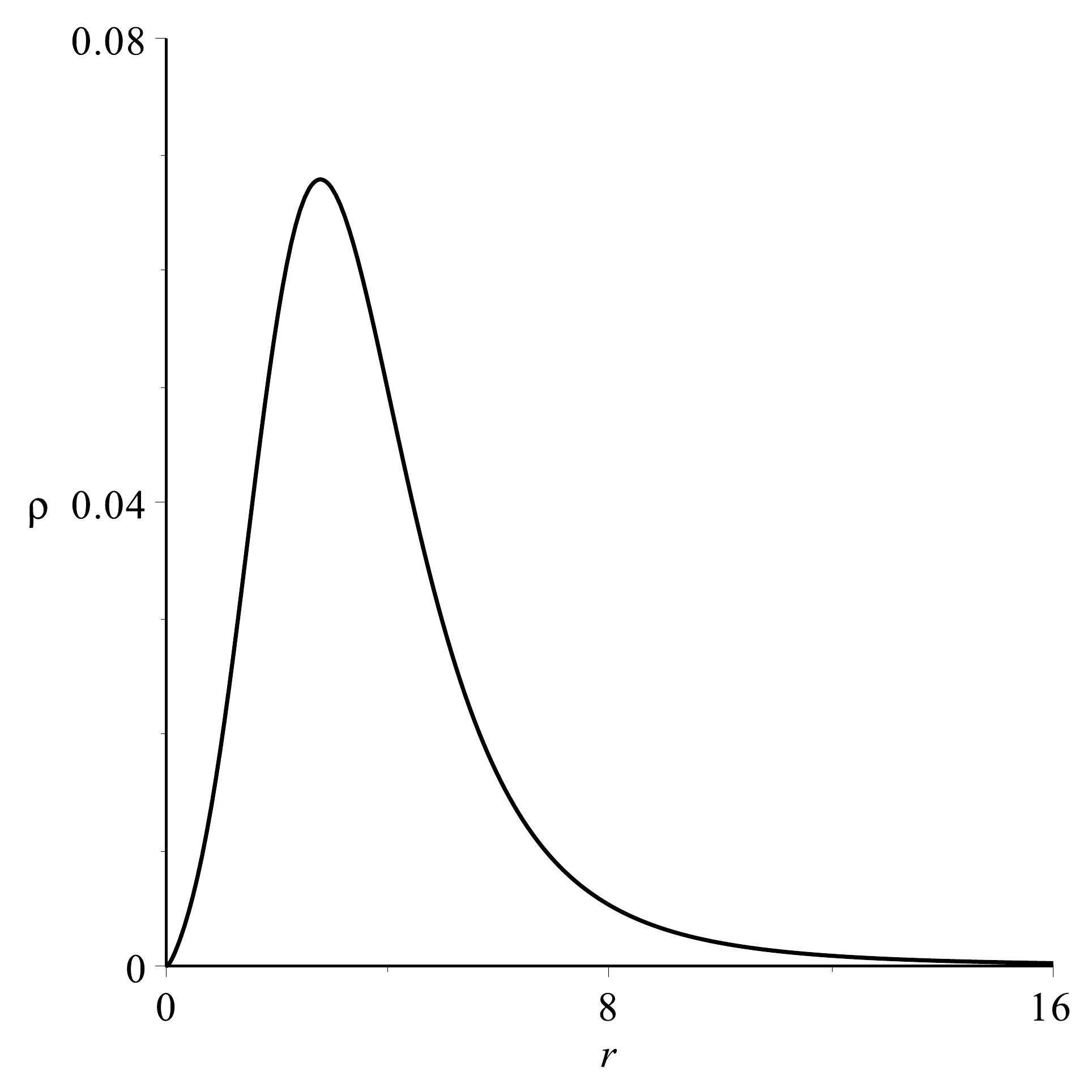}
\caption{The energy density \eqref{rhologcs}, displayed for the solutions $a(r)$ and $g(r)$ of Eqs.~\eqref{focslog} with $e,v,\kappa=1$.}
\label{figrhocs}
\end{figure} 

\begin{figure}[h]
\centering
\includegraphics[width=3.5cm]{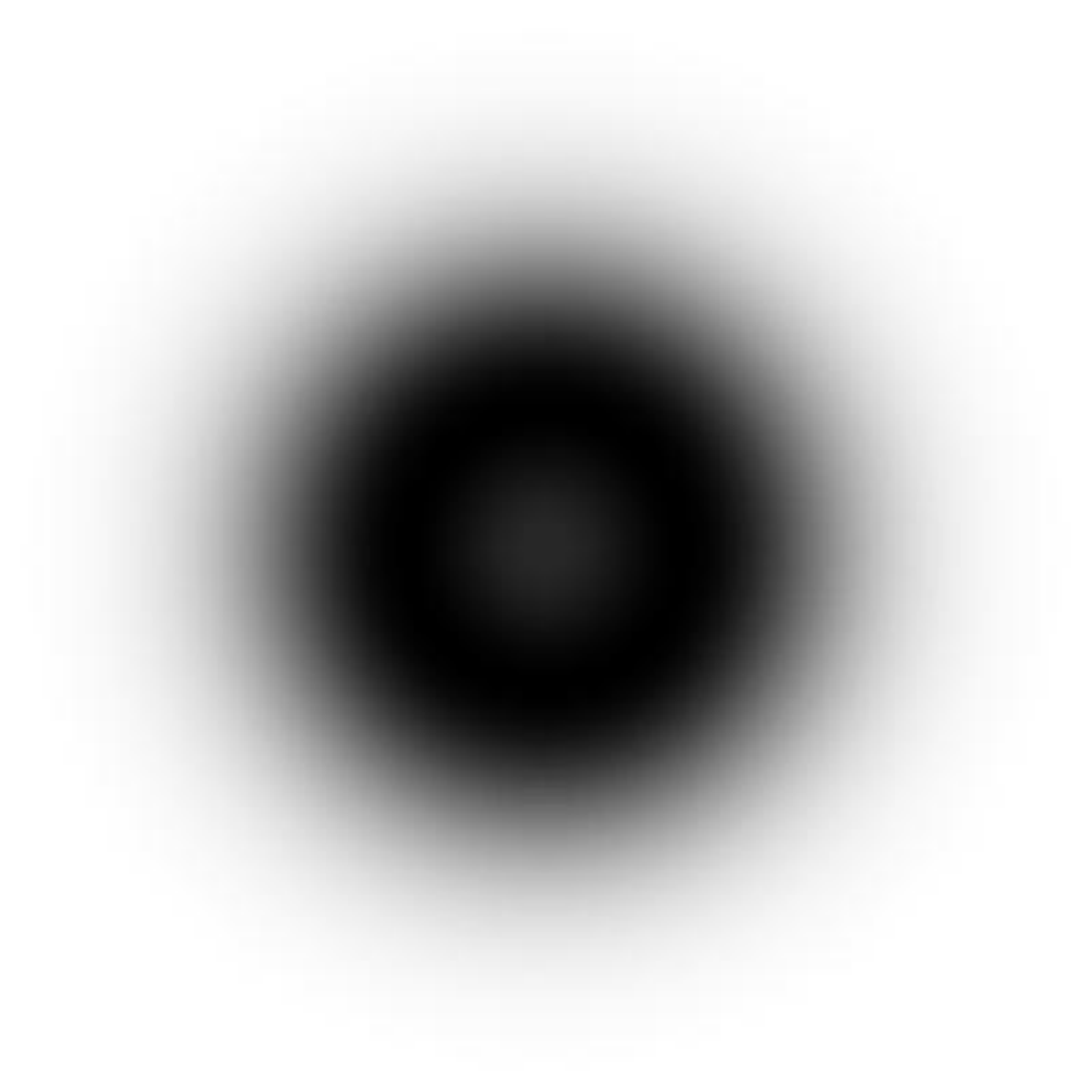}
\includegraphics[width=3.5cm]{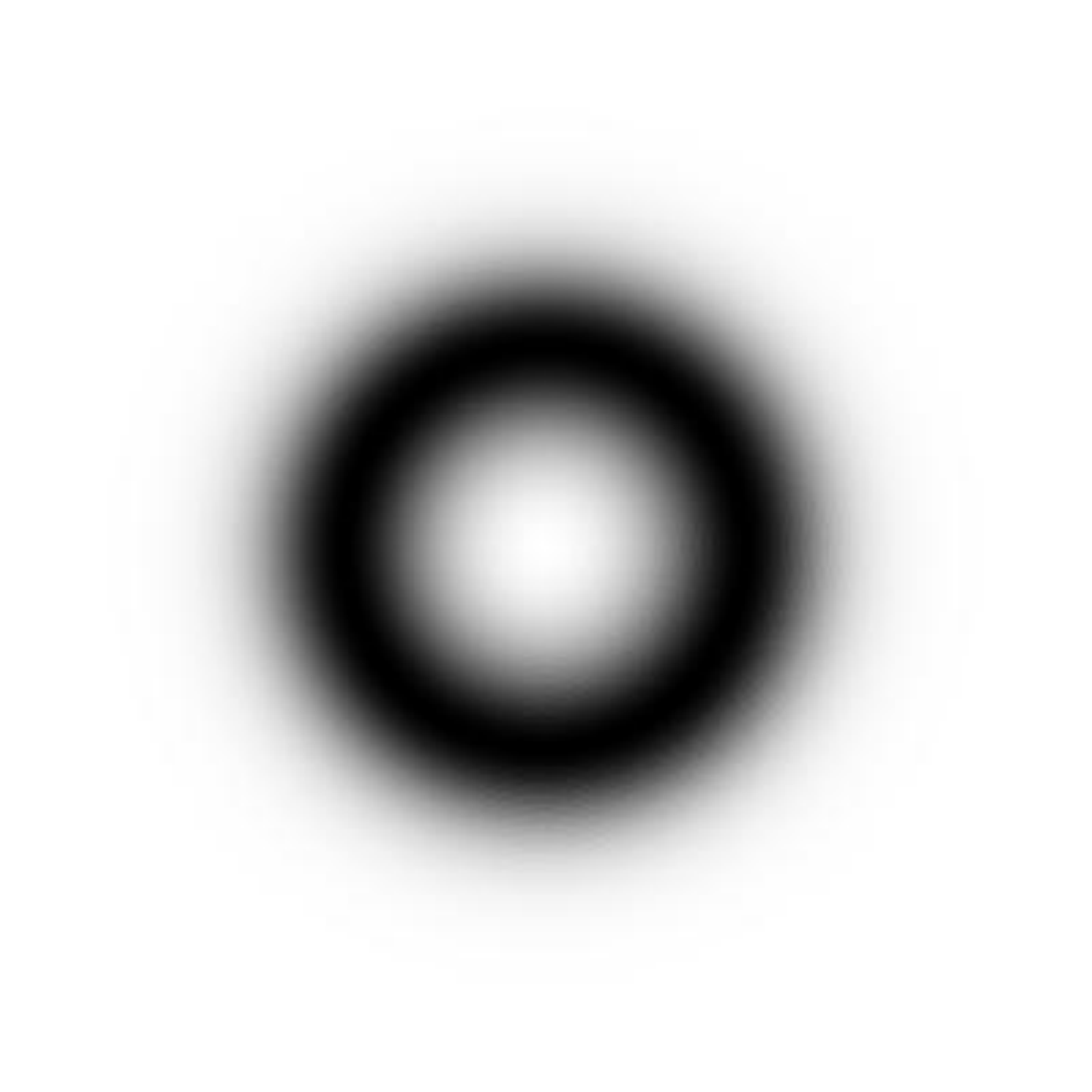}
\includegraphics[width=3.5cm]{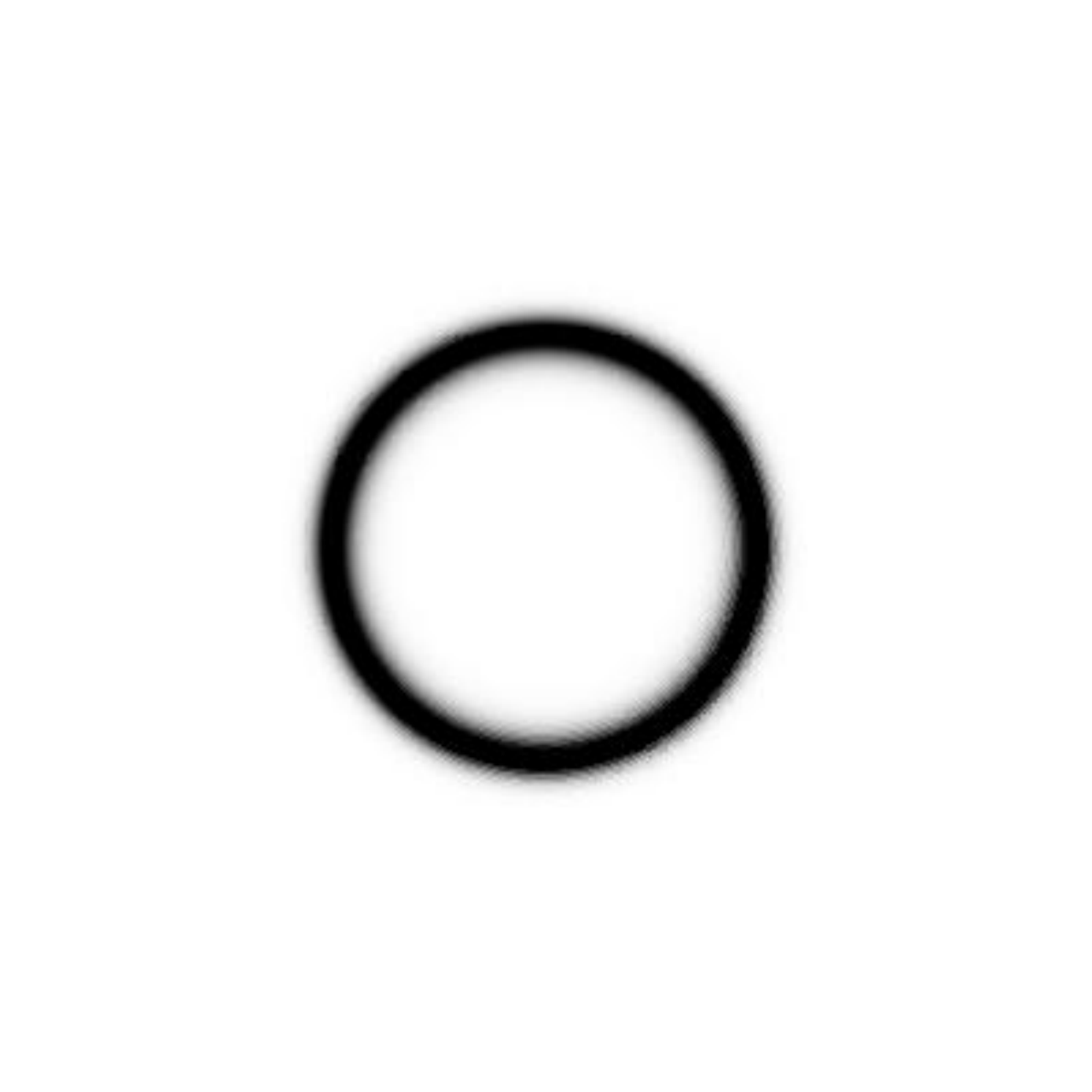}
\includegraphics[width=3.5cm]{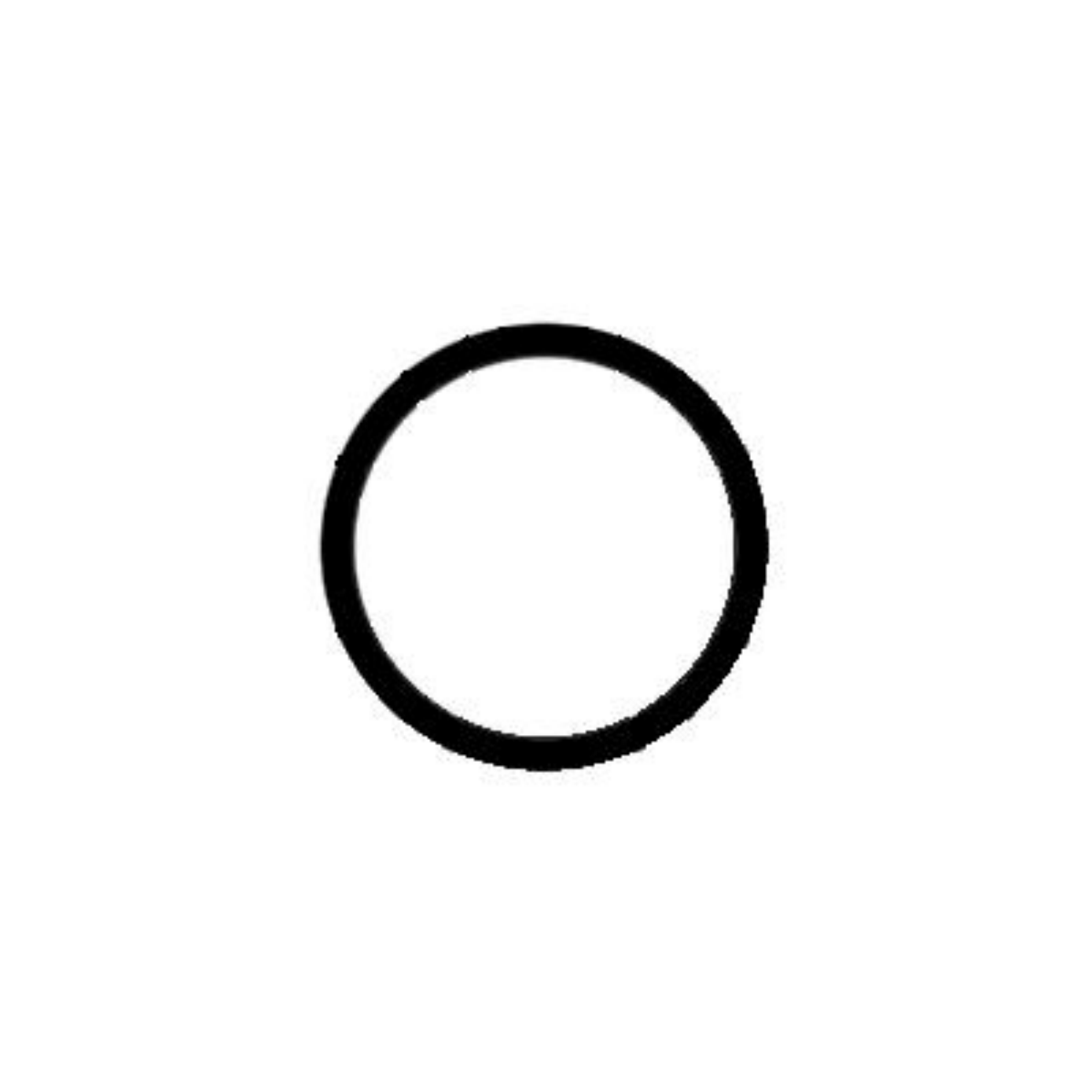}
\caption{The energy density associated to the vortex solution for the system defined by Eqs.~\eqref{kcs} and \eqref{vcs}, displayed in the $(r,\theta)$ plane for $e,v,\kappa,n=1$,  and for $l=1,2, 10,$ and $100$.}
\label{fig.cs}
\end{figure} 

Other examples of generalized models of the Chern-Simons type appeared recently in Ref.~\cite{CS}. In particular, the second model there investigated is very interesting since it gives rise to compact vortices of the Chern-Simons type, with the energy density (and also the electric and magnetic fields) being shrunk to a ringlike region in the $(r,\theta)$ plane. The model is of the type \eqref{gcsaction}, defined by
\begin{equation}\label{kcs}
K(|\varphi|)= l\,\left(\frac{|\varphi|}{v}\right)^{2l-2},
\end{equation}
and
\begin{equation}\label{vcs}
V(|\varphi|)=l\,\frac{e^4 v^6}{\kappa^2}\,\left(\frac{|\varphi|}{v}\right)^{2l}\left(1-
\left(\frac{|\varphi|}{v}\right)^{2l}\right)^2,
\end{equation}
where $l\geq1$ is a real number, with $l=1$ leading us back to the standard Chern-Simons model. Note that the above $K(|\varphi|)$ is exactly the same we have used before in Eq.~\eqref{illustm} for the Maxwell-Higgs system. In this case, it was shown in \cite{CS} that the vortex solution shrink to a ringlike region and here we illustrate this in Fig.~\ref{fig.cs}, for some values of $l$. We note that the the effect of increasing $l$ to make the vortex shrink to the compact region is much more significant than it appeared in the previous case, with the Maxwell-Higgs model there studied; see Fig.~\ref{fig5}. Also, the effect here is different since the solution shrinks to a hollow, ringlike region, not the disklike region that appeared in Fig.~\ref{fig5}. 

In Ref.~\cite{CS} the energy of the vortex was calculated numerically. Here, however, the formalism developed above allows to use the function $W$ which appears in \eqref{wcs} to obtain the total energy of the vortex configuration. We see that in terms of $a$ and $g$, the function $W$ can be written in the form
\begin{equation}
W=- a v^2 + a v^2 \left(\frac{g}{v}\right)^{2l}.
\end{equation}
It is such that the total energy becomes $E=2\pi|n|v^2$, and for $v=1$ and $n=1$ one has $E=2\pi$, in accordance with the numerical result obtained in \cite{CS}.

\section{Maxwell-Chern-Simons-Higgs Vortices}\label{mcs}

In this section, we consider a generalized action for the Maxwell-Chern-Simons system, of the form
\be\label{gmcsaction}
S=\int d^3x\left[\LL(X,Y,Z, |\vphi|,N) + \frac{\kappa}{4}\epsilon^{\alpha\beta\gamma}A_\alpha F_{\beta\gamma}\right].
\ee
Here, $X$ and $Y$ are given by Eq.~\eqref{XY} and $Z=\frac{1}{2}\partial_{\mu}N\partial_{\mu}N$, where $N$ is a neutral scalar field which we include to help to implement the first order formalism which we have obtained in the previous sections. The variation of the action \eqref{gmcsaction} with respect to the fields $\vphi$, $A_\mu$ and $N$ gives the equations of motion
\bes\label{gmcseom}
\begin{align}
 D_\mu (\LL_X D^\mu\vphi)&= \frac{\vphi}{2|\vphi|}\LL_{|\vphi|}, \\ \label{mcseqs}
 \partial_{\mu}(\LL_YF^{\mu\lambda})+\frac{\kappa}{2} \epsilon^{\lambda\mu\nu}F_{\mu\nu} &= J^\lambda, \\
 \partial_{\mu}(\LL_Z\partial^{\mu}N)&= \LL_N,
\end{align}
\ees
where the current is $J_\mu = ie{\cal L}_X(\bar{\vphi}D_\mu \vphi-\vphi\overline{D_\mu\vphi})$. It is possible to expand these equations to get
\bes\label{gmcseomexp}
\begin{align}
&\LX D_\mu D^\mu \vphi + 2\LL_{XX}D^\mu\vphi\,\Re\left(\overline{D_\alpha\vphi} \partial_\mu D^\alpha \vphi \right) -\frac12\LL_{XY}D^\mu\vphi F_{\alpha\beta}\partial_\mu F^{\alpha\beta} +\LL_{XZ} D^\mu\vphi \partial_\alpha N\partial_\mu\partial^\alpha N \nonumber \\
 &= \frac{\vphi}{2|\vphi|}\LL_{|\vphi|} -\LL_{X|\vphi|}D^\mu\vphi\,\Re\left(\frac{\overline{\vphi}}{|\vphi|}\partial_\mu\vphi\right) - \LL_{XN} D^\mu\vphi\partial_\mu N, \\
 &\LL_Y\partial_{\mu}F^{\mu\lambda} + 2\LL_{YX}F^{\mu\lambda}\,\Re\left(\overline{D_\alpha\vphi} \partial_\mu D^\alpha \vphi \right) -\frac12\LL_{YY}F^{\mu\lambda} F_{\alpha\beta}\partial_\mu F^{\alpha\beta} +\LL_{YZ} F^{\mu\lambda} \partial_\alpha N\partial_\mu\partial^\alpha N \nonumber\\
 &= J^\lambda -\frac{\kappa}{2} \epsilon^{\lambda\mu\nu}F_{\mu\nu} -\LL_{Y|\vphi|}F^{\mu\lambda} \,\Re\left(\frac{\overline{\vphi}}{|\vphi|}\partial_\mu\vphi\right) - \LL_{YN} F^{\mu\lambda}\partial_\mu N, \\
 &\left(\LL_Z\eta^{\mu\nu} +\LL_{ZZ}\partial^\mu N\partial^\nu N\right)\partial_\mu\partial_\nu N+2\LL_{ZX}\partial^\mu N\,\Re\left(\overline{D_\alpha\vphi} \partial_\mu  D^\alpha \vphi \right) -\frac12\LL_{ZY}\partial^\mu N F_{\alpha\beta}\partial_\mu F^{\alpha\beta}  \nonumber\\
 &= \LL_N -\LL_{Z|\vphi|}\partial^\mu N\,\Re\left(\frac{\overline{\vphi}}{|\vphi|}\partial_\mu\vphi\right) - \LL_{ZN} \partial^\mu N\partial_\mu N.
\end{align}
\ees
The energy momentum tensor $T_{\mu\nu}$ for the generalized model \eqref{gmcsaction} is given by 
\be
T_{\mu\nu}=\LL_YF_{\mu\lambda}{F^{\lambda}}_{\nu} +\LX\left( \overline{D_\mu \vphi}D_\nu \vphi + \overline{D_\nu \vphi}D_\mu \vphi\right) +\LL_Z(\partial_{\mu}N)(\partial_{\nu}N) - \eta_{\mu\nu} \LL.
\ee
Considering static solutions, the components of the energy-moment tensor have the form:
\bes\label{gTmcscomps}
\begin{align}\label{mgcsrho}
T_{00} &= \LL_Y E^2+2\LX e^2 A_0^2 |\vphi|^2-\LL, \\ 
T_{0i} &= -\LL_Y\epsilon_{ij}E^jB-A_0 J_i, \\
T_{12} &= -\LL_Y E_1E_2+\LX \left(  \overline{D_1 \vphi}D_2 \vphi + \overline{D_2 \vphi}D_1 \vphi\right) -\LL_Z\partial_{1}N\partial_{2}N, \\
T_{11} &= \LL_Y\left(B^2-(E_1)^2\right)+2\LX \left|D_1\vphi\right|^2 +\LL_Z(\partial_{1}N)^2 +\LL, \\ 
T_{22} &= \LL_Y\left(B^2-(E_2)^2\right)+ 2\LX \left|D_2\vphi\right|^2 + \LL_Z(\partial_{2}N)^2 + \LL.
\end{align}
\ees
In the above equations, we have used the definitions $E^i=(E_x,E_y) = F_{0i}$ and $B=-F^{12}$ for the electric and magnetic fields as in previous discussion just below the Eqs.~\eqref{gTcscompss}. However, for this new situation, the temporal component of Eq.~\eqref{mcseqs} gives that the magnetic field $B$ and the temporal component of the gauge field $A_0$ are related by $A_0 = \kappa B/(2e^2\LX|\vphi|^2 ) + (r\LL_Y A^\prime_0)^\prime/r$. Although this expression for $A_0$ has an additional term when compared to \eqref{A0}, one can show that the charge still can be written in terms of the magnetic flux \eqref{mflux} as $Q = -\kappa \Phi$, which makes the electric charge quantized. The equations of motion Eqs.~\eqref{gmcseom} with the ansatz \eqref{ansatz} and $A_0=A_0(r)$, with boundary conditions $A^\prime_0(0)=0$ and $A_0(\infty)=0$, are given by
\bes\label{eommcsansatz}
\begin{align}\label{eommcsansatzg}
\frac{1}{r} \left(r\LX g^\prime\right)^\prime + \LX g \left(e^2 A_0^2-\frac{a^2}{r^2} \right) + \frac12 \LL_{|\vphi|} &= 0,  \\ \label{amcsansatz}
r\left(\frac{\LL_Y a^\prime}{r}\right)^\prime - \kappa erA_0^\prime - 2e^2\LX a g^2 &= 0, \\ \label{a0mcsansatz}
 \frac{1}{r}(r\LL_Y A^\prime_0)^\prime-\frac{\kappa}{e}\frac{a^\prime}{r} - 2e^2\LX g^2 A_0 &= 0,\\\label{nmcsansatz}
 \frac{1}{r}\left(r\LL_Z N^\prime\right)^\prime + \LL_N &=0. 
\end{align}
\ees
Following Eqs.~\eqref{gmcseomexp}, we can expand the above equations into
\bes\label{eommcsexpansatz}
\begin{align}\label{eommcsexpansatzg}
\LX g^{\prime\prime} +  \left(\LL_X^\prime +\frac{\LX}{r}\right)g^\prime + \LX g \left(e^2 A_0^2-\frac{a^2}{r^2} \right) + \frac12 \LL_{|\vphi|} &= 0,  \\ 
\LY a^{\prime\prime} +\left(\LL_Y^\prime -\frac{\LL_Y}{r}\right)a^\prime - \kappa erA_0^\prime - 2e^2\LX a g^2 &= 0, \\ 
\LY A_0^{\prime\prime} +\left(\LL_Y^\prime +\frac{\LL_Y}{r}\right)A_0^\prime -\frac{\kappa}{e}\frac{a^\prime}{r} - 2e^2\LX g^2 A_0 &= 0,\\
\LL_Z N^{\prime\prime} +\left(\LL_Z^\prime +\frac{\LL_Z}{r}\right)N^\prime  + \LL_N &=0.
\end{align}
\ees
Since the prime means the derivative with respect to the radial coordinate $r$, one has $\LX^\prime = \LXX X^\prime + \LL_{XY}Y^\prime + \LL_{XZ}Z^\prime + \LL_{XN}N^\prime + \LL_{X|\vphi|}g^\prime$. The functions  $\LY^\prime$ and $\LL_{Z}^\prime$ are obtained similarly.

In this case, we have $E^i=-x^i A_0^\prime/r$, $B=-a^\prime/(er)$ and
\be
X=e^2g^2A_0^2-\left({g^\prime}^2+\frac{a^2g^2}{r^2}\right),\quad Y=\frac12{A_0^\prime}^2-\frac{{a^\prime}^2}{2e^2r^2} \quad\text{and}\quad Z=-\frac12{N^\prime}^2.
\ee
The components of the energy momentum tensor in Eq.~\eqref{gTmcscomps} become
\bes\label{gTmcscomp}
\begin{align}
T_{00} &= \LL_Y {A^\prime_0}^2 + 2\LL_X e^2g^2A^2_0 -\LL, \label{gTmcscomp0} \\
T_{01} &= \left(-\frac{\LL_Y A^\prime_0 a^\prime}{e}-2\LX e ag^2 A_0 \right)\frac{\sin{\theta}}{r},  \\
T_{02} &= \left(\frac{\LL_Y A^\prime_0 a^\prime}{e}+2\LX e ag^2 A_0 \right)\frac{\cos{\theta}}{r}, \\
T_{12} &= \left(2\LX \left( {g^\prime}^2 - \frac{a^2g^2}{r^2} \right) -\LL_Y {A^\prime_0}^2 +\LL_Z{N^\prime}^2\right)\frac{\sin(2\theta)}{2}, \\ 
T_{11} &=\LL_Y \frac{{a^\prime}^2}{e^2r^2} + 2\LX \left({g^\prime}^2\cos^2\theta+\frac{a^2g^2}{r^2}\sin^2\theta \right)-\left(\LL_Y {A^\prime_0}^2 - \LL_Z {N^\prime}^2\right)\cos^2\theta +\LL, \\
T_{22} &=\LL_Y \frac{{a^\prime}^2}{e^2r^2} + 2\LX \left({g^\prime}^2\sin^2\theta+\frac{a^2g^2}{r^2}\cos^2\theta \right) -\left(\LL_Y {A^\prime_0}^2 - \LL_Z {N^\prime}^2\right)\sin^2\theta +\LL.
\end{align}
\ees
For stressless solutions, we take $T_{12} = 0$, which gives
\be\label{foeqmcs1}
2\LX \left( {g^\prime}^2 - \frac{a^2g^2}{r^2} \right) -\LL_Y {A^\prime_0}^2 +\LL_Z{N^\prime}^2=0.
\ee
By using this, we get
\be
T_{11} = T_{22} = \LL_Y \frac{{a^\prime}^2}{e^2r^2} +2\LX {g^\prime}^2 -\LL_Y {A^\prime_0}^2 +\LL_Z{N^\prime}^2 + \LL.
\ee
Now, proceeding in the same way as in the previous sections we can rescale the energy to show that $T_{11}$=$T_{22}$=0, or
\be\label{foeqmcs2}
 \LL_Y \frac{{a^\prime}^2}{e^2r^2} +2\LX {g^\prime}^2 -\LL_Y {A^\prime_0}^2 +\LL_Z{N^\prime}^2 + \LL=0.
\ee
One can take the derivative of the the first order equation \eqref{foeqmcs2} and use \eqref{foeqmcs1} to get
\be\label{dderrickmcs}
\begin{aligned}
& 2g^\prime \left(\frac{1}{r} \left(r\LX g^\prime\right)^\prime + \LX g \left(e^2 A_0^2-\frac{a^2}{r^2} \right) + \frac12 \LL_{|\vphi|} \right) + \frac{a^\prime}{e^2r^2}\left( r\left(\frac{\LL_Y a^\prime}{r}\right)^\prime - \kappa erA_0^\prime - 2e^2\LX a g^2\right)\\
&-A_0^\prime \left(\frac{1}{r}(r\LL_Y A^\prime_0)^\prime-\frac{\kappa}{e}\frac{a^\prime}{r} - 2e^2\LX g^2 A_0 \right) + N^\prime\left(\frac{1}{r}\left(r\LL_Z N^\prime\right)^\prime + \LL_N\right)=0
\end{aligned}
\ee
which is a combination of the equations of motion \eqref{eomcsansatz}. In this case, even though we have four equations of motion given by \eqref{eommcsansatz}, we only have two first order equations, given by Eqs.~\eqref{foeqmcs1} and \eqref{foeqmcs2}. We then need to find more first order equations to solve the problem. Before doing so, we observe that, by using Eq.~\eqref{foeqmcs2}, the energy density \eqref{gTmcscomp0} can be written as
\be\label{rhomcs}
\rho = 2\LL_X({g^\prime}^2 + e^2g^2A^2_0) +  \LL_Y \frac{{a^\prime}^2}{e^2r^2} + \LL_Z{N^\prime}^2.
\ee
In order to get the correct first order equations in the standard case, $\LL=X+Y+Z-V(|\vphi|,N)$, which were obtained in Ref.~\cite{nmcs} through the BPS procedure, we suppose that Eq.~\eqref{godeq} holds, and combine it with Eq.~\eqref{foeqmcs1} and \eqref{foeqmcs2}, to get
\bes\label{fomcs}
\begin{align}
{g^\prime}^2 - \frac{a^2g^2}{r^2} &=0,\\ \label{mcsa0n}
\LL_Y {A^\prime_0}^2 -\LL_Z{N^\prime}^2&=0,\\ \label{fomcsderrick}
\LL+2\LX {g^\prime}^2+\frac{\LL_Y{a^\prime}^2}{e^2r^2} &= 0.
\end{align}
\ees
At this point, we see the importance of the neutral scalar field: it drives the temporal component of the gauge field; without it, the first order equation \eqref{mcsa0n} leads to a nonvanishing (due to the Chern-Simons term) constant $A_0$ and to a divergent energy as we can see from Eq.~\eqref{rhomcs}. Here, we remark that Eqs.~\eqref{fomcs} are the only possible first order equations. However, they are not enough to completely solve the problem, because we have to find the solution for the functions $a(r)$, $g(r)$, $N(r)$ and $A_0(r)$. Therefore, in the class \eqref{gmcsaction} of generalized Maxwell-Chern-Simons vortices, the use of one of the equations of motion \eqref{eommcsansatz} is required. By considering the Gauss' law \eqref{a0mcsansatz} for our model, it is possible to write the energy density as
\be
\rho = \frac{1}{r} \left(r\LL_Y A_0 A_0^\prime\right)^\prime - \LL_Y{A_0^\prime}^2 -\frac{\kappa A_0 a^\prime}{er} + 2\LL_X{g^\prime}^2 + \LL_Y \frac{{a^\prime}^2}{e^2r^2} + \LL_Z{N^\prime}^2.
\ee
We now use the first order equations \eqref{fomcs} and take an auxiliar function $W=W(a,g)$, such that
\be\label{wmcs}
W_a = \LL_Y\frac{a^\prime}{e^2r} - \frac{\kappa A_0}{e} \quad\text{and}\quad W_g=2r\LX g^\prime.
\ee
By doing that, the energy density can be written as a total derivative
\be
\rho = \frac{1}{r} \frac{d}{dr}\left[r\LL_Y A_0A_0^\prime + W(a,g) \right].
\ee
Since the first term vanishes when integrated all over the space, the energy is then given by
\be\label{energywmcs}
E = 2\pi \left|W\left(a(\infty),g(\infty)\right)-W\left(a(0),g(0)\right)\right|.
\ee
This formalism helps to calculate the energy without knowing the solutions. We omit further details here, since the issue is similar to the case studied before in Sec.~\ref{maxwell}.

To check the compatibility of the above first-order equations with the equations of motion \eqref{eommcsansatz}, we consider $g^\prime = ag/r$ in Eq.~\eqref{eommcsexpansatzg} to get
\be\label{mcsconst1}
\frac{a^\prime}{r} g\LX + \frac{ag}{r}\LX^\prime + e^2\LX g A_0^2+ \frac12\LL_{|\vphi|}=0,
\ee
where $\LX^\prime = \LXX X^\prime + \LL_{XY}Y^\prime + \LL_{XZ}Z^\prime + \LL_{XN}N^\prime + \LL_{X|\vphi|}g^\prime$. By reminding that Eq.~\eqref{mcsconst1} comes from the equation of motion \eqref{eommcsansatzg}, one can show that, if the first order equations \eqref{fomcs} are compatible with two of the three other equations of motion, from Eq.~\eqref{dderrickmcs} we conclude that all of the equations of motion \eqref{eommcsansatz} are satisfied. This means that one of the equations of motion is an identity. In this point, proceeding similarly as in the previous sections, if we want to construct the model with analytical Lagrangian densities that support the proceedure \eqref{energywmcs}, we can take the path of setting $\LXX=\LL_{XY}=\LL_{XZ}=\LL_{XN}=0$ and $\LL_{X|\vphi|}=K_{|\vphi|}(|\vphi|)$ to get a constraint that only depends on the functions $g$ and $N$, which represent the scalar fields. Since we are dealing with two scalar fields in the model, a second constraint is needed to completely determine how the scalar fields must appear in the Lagrangian density. To search for it, we see that Eqs.~\eqref{a0mcsansatz} and \eqref{nmcsansatz} produces 
\be\label{mcsconst2}
\frac{1}{r} \left[r\!\left(\LL_Z N^\prime - \LL_Y A_0^\prime\right) \right]^\prime + \frac{\kappa a^\prime}{er} + 2e^2 \LX g^2 A_0 + \LL_N=0.
\ee
We now make use of the first order equation \eqref{mcsa0n} and take $\LL_Y=\LL_Z$ to see that, by using $A_0^\prime=N^\prime$, the above equation constrains the field $N$ in the Lagrangian density. In order to get a class of models that can be constructed analytically, we consider
\be\label{mcsbestl}
\LL = K(|\vphi|) X + G(U,|\vphi|,N),\quad\text{where}\quad U = Y + Z.
\ee
For this Lagrangian density, the first order equation \eqref{mcsa0n} reads ${A_0^\prime}^2 = {N^\prime}^2$. We then take $A_0(r) = N(r)$. In this case, we have $U=-{a^\prime}^2/(2e^2r^2)$ and Eq.~\eqref{fomcsderrick} gives
\be\label{derrickmcsgu}
e^2g^2N^2K + G-2UG_U=0.
\ee
The above equation is an algebraic equation that relates $U$, $g$ and $N$. Considering that it can be solved for $U$, we write
\be\label{mcsveff}
-U = \frac{{a^\prime}^2}{2e^2r^2} = V_{eff}(g,N).
\ee
We then take $a^\prime/(er) = -\sqrt{2V_{eff}}$ and combine it with Eqs.~\eqref{mcsconst1} and \eqref{mcsconst2} to get the constraints
\bes\label{mcsconstraints}
\begin{align}
egK\sqrt{2V_{eff}} & = e^2N^2gK + \frac{e^2}{2}N^2g^2K_{|\vphi|} + \frac12 G_{|\vphi|}, \\
\kappa\sqrt{2V_{eff}} &= 2e^2Ng^2K+G_N.
\end{align}
\ees
Above, we have two partial differential equations that completely determines how the Lagrangian density depends on the fields $\vphi$ and $N$. In this case, if the constraints \eqref{mcsconstraints} are satisfied, we can use \eqref{wmcs} to get
\be
W(a,g) = -\frac{a}{e}\left(G_U\sqrt{-2U}+\kappa N \right)_{-U=V_{eff}}.
\ee
Thus, the energy can also be calculated analytically.

We now consider the model \eqref{mcsbestl} with a general $K(|\vphi|)$ and
\be
G(U,|\vphi|,N) = -H(|\vphi|,N)(-U)^s - V(|\vphi|,N).
\ee
Here, $s$ is a real parameter such that $s>1/2$. The case $s=1$ was considered in Ref.~\cite{gmcsbazeia}. The standard case is obtained for $K(|\vphi|)=H(|\vphi|,N)=s=1$ and $s=1$. We use Eqs.~\eqref{derrickmcsgu} and \eqref{mcsveff} for the above function to get the effective potential
\be
V_{eff}(|\vphi|,N) = \left(\frac{V(|\vphi|,N)-e^2N^2|\vphi|^2K(|\vphi|)}{(2s-1)H(|\vphi|,N)}\right)^{1/s}.
\ee
The constraints in Eqs.~\eqref{mcsconstraints} become
\bes\label{const1mcs}
\bal
\left(2V_{eff}\frac{\partial H}{\partial g} +(2s-1)H\frac{\partial V_{eff}}{\partial g}\right)sV_{eff}^{s-1} &= -2egK\sqrt{2V_{eff}}, \\
\left(2V_{eff}\frac{\partial H}{\partial N} +(2s-1)H\frac{\partial V_{eff}}{\partial N}\right)sV_{eff}^{s-1} &= -\kappa\sqrt{2V_{eff}}.
\eal
\ees
The solution of the above equations leads to the potential
\be\label{potmcs1}
V(|\vphi|,N) = (2s-1)H(|\vphi|,N)\left(\frac{ev^2 - e\int_0^{|\vphi|}d\tilde{g}\, 2\,\tilde{g}\,K(\tilde{g}) - \kappa N}{s\sqrt{2}H(|\vphi|,N)}\right)^{\frac{2s}{2s-1}} + e^2N^2|\vphi|^2 K(|\vphi|),
\ee
where $v$ is a parameter that breaks the symmetry and $K(|\vphi|)$ is a function that leads to symmetry breaking of the potential. As in the previous scenarios, the above potential was taken because the standard model is straightforwardly obtained for $K(|\vphi|)=H(|\vphi|,N)=s=1$, leading to the potential $V_{std}(|\vphi|,N) = \left(ev^2-e|\vphi|^2-\kappa N \right)^{2}\!/2 + e^2N^2|\vphi|^2$.

The equations to be solved for the potential \eqref{potmcs1} with a general $K(|\vphi|)$ and $A_0(r) = N(r)$ are the first-order ones from Eqs.~\eqref{fomcs}
\be
{g^\prime} = \frac{ag}{r}, \quad\text{and}\quad a^\prime = -er\left(\frac{ev^2 - e\int_0^{g}d\tilde{g}\, 2\,\tilde{g}\,K(\tilde{g}) - \kappa N}{s\,2^{1-s}H(g,N)}\right)^{\frac{1}{2s-1}}.
\ee
As stated before, since we have four fields but only three first order equations, the use of one of the equations of motion \eqref{eommcsansatz} is required to solve the problem. In particular, we can use
\be
\frac{1}{r}(rN^\prime)^\prime = -\kappa\left(\frac{ev^2 - e\int_0^{g}d\tilde{g}\, 2\,\tilde{g}\,K(\tilde{g}) - \kappa N}{s\,2^{1-s}H(g,N)}\right)^{\frac{1}{2s-1}}+ 2e^2Ng^2 K(g)
\ee
In this case, to calculate the energy analitycally, we can use Eq.~\eqref{wmcs} to get
\be
W(a,g)=a\left(\int_0^{g}d\tilde{g}\, 2\,\tilde{g}\,K(\tilde{g})-v^2\right).
\ee
Considering that $W(0,v)=0$, we have $E= 2\pi|n|v^2$. As in the previous cases, this is a consequence of the potential taken in Eq.~\eqref{potmcs1}. Other energies may be obtained for different solutions of the constraint \eqref{const1mcs}.

It is straightforward to show that this result is also valid for the standard case, $H(|\vphi|,N)=1$, $K(|\vphi|)=1$ and $s=1$, in which we have $W(a,g) = a(g^2-v^2)$. Another model that falls into the class of systems that we have just obtained was studied numerically in \cite{gmcsbazeia}, but it does not lead to compact solutions.

\section{Comments and Conclusions}\label{conclusions}

In this work, we studied generalized Maxwell-Higgs, Chern-Simons-Higgs and Maxwell-Chern-Simons-Higgs models, which support vortex configurations. We investigated the existence of a first order formalism for the most general class of models that is possible in each one of the scenarios. As shown above, each of the three distinct cases must satisfy specific constraints. We have also introduced a method to calculate the energy without knowing the explicit solutions for any generalized model that satisfies the constraints.

In the Maxwell-Higgs scenario, there are two equations of motion, since the Gauss' law is an identity for uncharged vortex solutions. In this case, it is possible to obtain two first order equations, which appears motivated by rescaling arguments in the stress tensor. Nevertheless, we have shown that not every Lagrangian density supports stressless solutions. The class of models that do the job is restricted by a constraint that is very complicated. Then, we have introduced a path to construct the models analytically and calculated the auxiliar function $W=W(a,g)$ that allows to evaluate the energy without knowing the explicit form of the solutions. 

The Chern-Simons framework presents a slight difference: the Chern-Simons term in the Lagrangian density cannot be generalized because it is not gauge invariant. For the class of models that we studied, we have three equations of motion but only two first order equations to produce stressless solutions. For this reason, we used the Gauss' law as a third first order equation. Similarly as in the Maxwell-Higgs case, it was also possible to unveil a path to construct the Lagrangian density analytically and to calculate the auxiliary function $W=W(a,g)$ that allows to evaluate the energy of the stressless solutions. For Maxwell-Chern-Simons-Higgs models, the formalism is trickier; the addition of a neutral scalar field in the Lagrangian density is needed in order to get a first order formalism. In this case, we have four equations of motion, all of them of second order, and only three first order equations that appear from the stressless condition. Therefore, there is no other first order equation to completely solve the problem. Due to that, we have used the Gauss' law as our fourth differential equation. In order to construct the Lagrangian density analytically, we noticed that it is possible to follow the same steps we have implemented in the previous scenarios, but now with two constraints. The existence of an auxiliary function $W=W(a,g)$ to calculate the energy was also shown in this case.

It is important to emphasize here that in all the three systems, in the Maxwell-Higgs, in the Chern-Simons-Higgs and in the Maxwell-Chern-Simons-Higgs  cases, the equations of motion and the constraints that appear in order for the systems to obey first order equations also lead to the construction of the auxiliary function $W=W(a,g)$, from which one can calculate the energy of the field configurations exactly, without the need of the explicit form of the solutions themselves. As far as we can see, these results were not present in the vortex literature until now. 

The main results unveiled interesting ways to construct generalized models that satisfy specific constraints that allow the calculation of the energy without knowing the stressless solutions themselves. In particular, the results on compact solutions in Maxwell-Higgs and in Chern-Simons-Higgs models are of current interest, and further study in the Maxwell-Chern-Simons-Higgs system has to be implemented to find compact solutions. Other perspectives include the extension of the present formalism to the case of Abelian non-topological vortices, non-Abelian vortices and monopoles. One may also try to develop a similar procedure for models with the symmetry enlarged to $U(1)\times U(1)$, which is of interest in the study of superconducting strings \cite{witten} and also, to describe the inclusion of the so-called hidden sector \cite{hidden1,hidden2,hidden3}. Another line of investigation that would enlarge the scope of the current work should consider possible extensions of the models to the supersymmetric context, following the lines of Refs.~\cite{schap1,schap2}. These problems are currently under consideration, and we hope to report on them in the near future.

\acknowledgments{We would like to acknowledge the Brazilian agency CNPq for partial financial support. DB thanks support from grant 306614/2014-6, LL thanks support from grant 303824/2017-4, MAM thanks support from grant 140735/2015-1 and RM thanks support from grant 306826/2015-1.}


\end{document}